\documentclass[a4paper,11pt]{article}
\pdfoutput=1
\usepackage{jheppub}
\usepackage{caption,subcaption}
\usepackage{graphicx} % demo is just for the example
\usepackage{rotating,float,afterpage}
\usepackage{calrsfs}
\usepackage{slashed,cancel}
\DeclareMathAlphabet{\pazocal}{OMS}{zplm}{m}{n}
\RequirePackage[T1]{fontenc}
\usepackage[utf8x]{inputenc}
\usepackage{amsmath,graphicx,epsfig,amssymb,multirow}
\RequirePackage{mathptmx,flushend}      % use Times fonts if available on your TeX system
\usepackage[12hr]{datetime} %to be removed finally
\allowdisplaybreaks

\def\wtil#1{\widetilde{#1}}
\def\MGvATNLO{{\tt {\sc MadGraph5}\_aMC@NLO}}
\def\l{\left}
\def\r{\right}
\definecolor{greena}{rgb}{0.13, 0.55, 0.13}
\title{Unravelling the anomalous gauge boson couplings in  $ZW^\pm$ production at the LHC  and the role of  spin-$1$  polarizations}
\author{Rafiqul Rahaman and} 
\author{Ritesh K. Singh}
\affiliation{Department of Physical Sciences,
    Indian Institute of Science Education and Research Kolkata,
    Mohanpur, 741246, India}
\emailAdd{rr13rs033@iiserkol.ac.in}
\emailAdd{ritesh.singh@iiserkol.ac.in} 
\abstract{
We study the anomalous triple gauge boson couplings (aTGC) in $ZW^\pm$ production in $3l+\cancel{E}_T$ channel at the LHC
for $\sqrt{s}=13$ TeV. We use cross sections, azimuthal asymmetry, forward-backward asymmetry, 
and polarization asymmetries of $Z$ and  reconstructed $W$ to estimate simultaneous limits
on the anomalous  couplings for both the effective vertex formalism as well 
as the effective operator approach using the Markov-chain–Monte-Carlo (MCMC) method for luminosities 
$35.9$ fb$^{-1}$, $100$ fb$^{-1}$, $300$ fb$^{-1}$, $1000$ fb$^{-1}$, and $3000$ fb$^{-1}$. 
The trilepton invariant mass ($m_{3l}$)
and the transverse momentum of $Z$ ($p_T(Z)$) are found to be sensitive to the aTGC for the cross sections
as well as for the asymmetries. We observed that the asymmetries significantly improve the
measurement of anomalous couplings at the high-luminosity large hadron collider (LHC) if a deviation from the Standard Model (SM) is observed.
}
\begin{document}
    \maketitle
%%%%%%%%%%%%%%%%%

%%%%%%%%%%%%%%%%%%%%%%%%%%%%%%%%%%%%%%%%%%%%%%%%%%%%%%%%%%%%%%%%%%%%%%%%%%%%%%%%%%%%%%%%%%%%%%%%%
\section{Introduction}
After the discovery of Higgs~\cite{Chatrchyan:2012xdj,Aad:2012tfa}, the Large Hadron Collider (LHC) has been looking for
new physics beyond the SM (BSM) needed to address many open questions such as 
neutrino oscillation, dark matter, baryogenesis, etc. with higher energies and
higher luminosities. Unfortunately, no new physics has been found~\cite{Summary:ATLAS:exotics,Summary:ATLAS:SUSY,Summary:CMS:exotic,Summary:CMS:longlived,Summary:CMS:B2G}  except a few 
fluctuations (e.g., Refs.~\cite{Aaboud:2016tru,Khachatryan:2016hje,Sirunyan:2018wim}). One could expect 
that the new physics scale is too heavy  to be directly 
explored by the LHC, and they may leave some footprints in the  available energy range. 
They will modify the structure of the SM vertices or bring some new vertices, 
often through higher-dimensional operators with the SM fields.
These new vertices and/or the  extra contribution to the SM  vertices are 
termed as anomalous in the sense that they are not present in the SM at leading order (LO). 
The electroweak sector will get affected by the anomalous bosonic self-couplings, which alter the paradigm of electroweak symmetry breaking (EWSB). To understand the EWSB mechanism, one needs precise measurements of the  couplings in the bosonic sector of the SM. 
 Here, we choose to focus on the charge sector by probing the $WWZ$ anomalous couplings in the $ZW^\pm$ production at the LHC. 
The $WWZ$ anomalous triple gauge boson couplings (aTGC) may be obtained by higher dimension effective 
operators made out of SM fields suppressed  by a new physics scale $\Lambda$. The effective 
Lagrangian including the higher dimension effective operators (${\cal O}$) 
 to the SM Lagrangian (${\cal L}_{SM}$) is treated to be
\begin{equation}\label{eq:eft-Lag}
    {\cal L}_{eft} = {\cal L}_{SM} + \sum_i \frac{c_i^{{\cal O}(6)}}{\Lambda^2}{\cal O}_i^{(6)}
    + \sum_i \frac{c_i^{{\cal O}(8)}}{\Lambda^4}{\cal O}_i^{(8)}  + \dots \ \ ,
\end{equation}
with $c_i^{{\cal O}(6,8)}$ being the couplings of the dimension-$(6,8)$  operators ${\cal O}_i^{(6,8)}$.
The effective operators n the Hagiwara-Ishihara-
Szalapski-Zeppenfeld (HISZ) basis up to dimension-$6$ contributing to  $WWZ/\gamma$   couplings,  in general, 
are~\cite{Hagiwara:1993ck,Degrande:2012wf} 
\begin{eqnarray}\label{eq:dim6-operators}
{\cal O}_{WWW}&=&\mbox{Tr}[W_{\mu\nu}W^{\nu\rho}W_{\rho}^{\mu}] \ \ ,\nonumber\\
{\cal O}_W&=&(D_\mu\Phi)^\dagger W^{\mu\nu}(D_\nu\Phi) \ \ ,\nonumber\\
{\cal O}_B&=&(D_\mu\Phi)^\dagger B^{\mu\nu}(D_\nu\Phi) \ \ ,\nonumber\\
{\cal O}_{\wtil{WWW}}&=&\mbox{Tr}[{\tilde W}_{\mu\nu}W^{\nu\rho}W_{\rho}^{\mu}] \ \ ,\nonumber\\
{\cal O}_{\wtil W}&=&(D_\mu\Phi)^\dagger {\tilde W}^{\mu\nu}(D_\nu\Phi) \ \ .
\end{eqnarray}
Among these operators ${\cal O}_{WWW}$,  ${\cal O}_W$ and  ${\cal O}_B$ are
$CP$-even, while ${\cal O}_{\wtil{WWW}}$  and ${\cal O}_{\wtil W}$ are
$CP$-odd. On the other hand, the $WWZ$ anomalous couplings may be parametrized
in a model independent way with the most general Lorentz invariant  
form factors or vertex factors given by~\cite{Hagiwara:1986vm}
\begin{eqnarray}
{\cal L}_{WWZ} &=&ig_{WWZ}\Big[\left(1+\Delta g_1^Z\right)(W_{\mu\nu}^+W^{-\mu}-
W^{+\mu}W_{\mu\nu}^-)Z^\nu
+\frac{\lambda^Z}{m_W^2}W_\mu^{+\nu}W_\nu^{-\rho}Z_\rho^{\mu}\nonumber\\
&+&\frac{\wtil{\lambda^Z}}{m_W^2}W_\mu^{+\nu}W_\nu^{-\rho}\wtil{Z}_\rho^{\mu}
+\left(1+\Delta \kappa^Z\right) W_\mu^+W_\nu^-Z^{\mu\nu}+\wtil{\kappa^Z}W_\mu^+W_\nu^-\wtil{Z}^{\mu\nu}
\Big] \ \ ,
\label{eq:Lagrangian}
\end{eqnarray}
where $W_{\mu\nu}^\pm = \partial_\mu W_\nu^\pm - 
\partial_\nu W_\mu^\pm$, $Z_{\mu\nu} = \partial_\mu Z_\nu - 
\partial_\nu Z_\mu$, 
$\wtil{Z}^{\mu\nu}=1/2\epsilon^{\mu\nu\rho\sigma}Z_{\rho\sigma}$,
and the overall coupling constants is given as
 $g_{WWZ}=-g\cos\theta_W$, $\theta_W$ being the weak
mixing angle.  The couplings $\Delta g_1^Z$, $\Delta\kappa^Z$
and $\lambda^Z$ of Eq.~(\ref{eq:Lagrangian}) are $CP$-even,  while  $\wtil{\kappa^Z}$
and $\wtil{\lambda^Z}$  are $CP$-odd in nature.  All the anomalous couplings vanish in the SM.
In the $SU(2)\times U(1)$ gauge group, the coupling 
($c_i^{\cal L}$) of the Lagrangian in Eq.~(\ref{eq:Lagrangian}) are related
to  the couplings $c_i^{\cal O}$ of the operators in Eq.~(\ref{eq:dim6-operators}) 
as~\cite{Hagiwara:1993ck,Wudka:1994ny,Degrande:2012wf}
\begin{eqnarray}
\Delta g_1^Z & = & c_W\frac{m_Z^2}{2\Lambda^2} \ \ ,\nonumber\\
\lambda^Z &=& c_{WWW}\frac{3g^2m_W^2}{2\Lambda^2} \ \ ,\nonumber\\
 \wtil{\lambda^Z}&=&  c_{\wtil{WWW}}\frac{3g^2m_W^2}{2\Lambda^2} \ \ ,\nonumber\\
%\Delta\kappa^\gamma & = & (c_W+c_B)\frac{m_W^2}{2\Lambda^2} \ \ ,\nonumber\\
\Delta\kappa^Z & = & (c_W-c_B\tan^2\theta_W)\frac{m_W^2}{2\Lambda^2} \ \ ,\nonumber\\
%\wtil{\kappa^\gamma} & = & c_{\wtil{W}}\frac{m_W^2}{2\Lambda^2} \ \ ,\nonumber\\
\wtil{\kappa^Z} & = &
-c_{\wtil{W}}\tan^2\theta_W\frac{m_W^2}{2\Lambda^2} \ \ .
\label{eq:Operator-to-Lagrangian}
\end{eqnarray}
We label the anomalous couplings of three scenarios for later uses as follows:
The couplings of the operators in Eq.~(\ref{eq:dim6-operators}), the couplings of effective vertices
in ${\cal L}_{WWV}$ in Eq.~(\ref{eq:Lagrangian}) and the vertex couplings translated from the 
operators in Eq.~(\ref{eq:Operator-to-Lagrangian}) are labelled as $c_i^{\cal O}$, $c_i^{\cal L}$,
 and $c_i^{{\cal L}_g}$, respectively.

These anomalous gauge boson self-couplings may be obtained from some high scale new 
physics such as MSSM~\cite{Lahanas:1994dv,Arhrib:1996rj,Argyres:1995ib}, extra dimension~\cite{FloresTlalpa:2010rm,Lopez-Osorio:2013xka}, Georgi-Machacek model~\cite{Arroyo-Urena:2016gjt}, etc.
by integrating out the heavy degrees of freedom.
Some of these couplings can also be obtained at loop level within the SM~\cite{Argyres:1992vv,Papavassiliou:1993ex}. 
 There have been a lot of studies to probe the 
anomalous $WWZ/\gamma$ couplings  in the effective operator  method as well
as in the effective vertex factor approach  subjected to  $SU(2)\times U(1)$ invariance for various colliders:  for $e^+$-$e^-$ linear 
collider~\cite{Gaemers:1978hg,Hagiwara:1986vm,Bilchak:1984ur,Hagiwara:1992eh,
    Wells:2015eba,Buchalla:2013wpa,Zhang:2016zsp,Berthier:2016tkq,Bian:2015zha,Bian:2016umx,
    Choudhury:1996ni,Choudhury:1999fz,Rahaman:2019mnz}, for the Large Hadron electron collider (LHeC)
~\cite{Biswal:2014oaa,Cakir:2014swa,Li:2017kfk}, $e$-$\gamma$ collider~\cite{Kumar:2015lna} and for
 the LHC~\cite{Baur:1987mt,Dixon:1999di,Falkowski:2016cxu,Azatov:2017kzw,Azatov:2019xxn,Bian:2015zha,Campanario:2016jbu,Bian:2016umx,Butter:2016cvz,Biekotter:2018rhp,Baglio:2017bfe,Li:2017esm,Bhatia:2018ndx,Chiesa:2018lcs,Baglio:2018bkm,Baglio:2019uty}. 
Some $CP$-odd $WWV$ couplings have been studied in Refs.~\cite{Choudhury:1999fz,Li:2017esm}. 
Direct measurement of these charged aTGC has been performed  at the LEP~\cite{Abbiendi:2000ei,Abbiendi:2003mk,
    Abdallah:2008sf,Schael:2013ita}, Tevatron~\cite{Aaltonen:2007sd,Abazov:2012ze},  LHC~\cite{Aaboud:2017cgf,
    Sirunyan:2017bey,Aaboud:2017fye,Khachatryan:2016poo,
    Aad:2016ett,Aad:2016wpd,Chatrchyan:2013yaa,
    RebelloTeles:2013kdy,ATLAS:2012mec,Chatrchyan:2012bd,Aad:2013izg,
    Chatrchyan:2013fya,Sirunyan:2017jej,Sirunyan:2019gkh,Sirunyan:2019dyi,Sirunyan:2019bez} and Tevatron-LHC~\cite{Corbett:2013pja}. The most stringent constraints on the operators ($c_i^{\cal O}$) are obtained in 
   Ref.~\cite{Sirunyan:2019gkh} for $CP$-even
   ones and in Ref.~\cite{Aaboud:2017fye} for $CP$-odd ones, and they are listed in Table~\ref{tab:aTGC_constrain_form_collider}. 
These limits translated to  the effective vertices ($c_i^{{\cal L}_g}$) are also 
given in Table~\ref{tab:aTGC_constrain_form_collider}.
\begin{table*}[!ht]
    \centering
    \caption{\label{tab:aTGC_constrain_form_collider} The list of tightest constraints observed on the
  effective operators   and the effective vertices in $SU(2)\times U(1)$ gauge at $95\%$ C.L.  from experiments.}
    \renewcommand{\arraystretch}{1.5}
    \begin{tabular*}{\textwidth}{@{\extracolsep{\fill}}lll@{}}\hline
        $c_i^{\cal O}$            & Limits (TeV$^{-2}$)   & Remark\\\hline 
        $\frac{c_{WWW}}{\Lambda^2}$                    & $[-1.58,+1.59]$ &CMS $\sqrt{s}=13$ TeV, ${\cal L}=35.9$ fb$^{-1}$, $SU(2)\times U(1)$~\cite{Sirunyan:2019gkh} \\
        $\frac{c_{W}}{\Lambda^2}$                     & $[-2.00,+2.65]$ &CMS~\cite{Sirunyan:2019gkh} \\
        $\frac{c_{B}}{\Lambda^2}$                    & $[-8.78,+8.54]$ &CMS~\cite{Sirunyan:2019gkh} \\    
        $ \frac{c_{\widetilde{WWW}}}{\Lambda^2}$    &$[-11,+11]$  &ATLAS $\sqrt{s}=7(8)$ TeV, ${\cal L}=4.7(20.2)$ fb$^{-1}$ ~\cite{Aaboud:2017fye}\\
        $ \frac{c_{\widetilde{W}}}{\Lambda^2}$      &$[-580,580]$  &ATLAS~\cite{Aaboud:2017fye} \\
        \hline
        $c_i^{{\cal L}_g}$ & Limits ($\times 10^{-2}$) & Remark\\ \hline
        $\lambda^Z$ &  $[-0.65,+0.66]$ &CMS~\cite{Sirunyan:2019gkh}\\
        %$\Delta\kappa^\gamma$ &$[-4.4,+6.3]$&CMS $\sqrt{s}=8$ TeV, ${\cal L}=19$ fb$^{-1}$, $SU(2)\times U(1)$~\cite{Sirunyan:2017bey}\\
        $\Delta g_1^Z$ & $[-0.61,+0.74]$ &  CMS~\cite{Sirunyan:2019gkh}\\
        $\Delta\kappa^Z$ & $[-0.79,+0.82]$ &CMS~\cite{Sirunyan:2019gkh}\\
        $\wtil{\lambda^Z}$ & $[-4.7,+4.6]$ &ATLAS~\cite{Aaboud:2017fye}\\
        $\wtil{\kappa^Z}$  & $[-14,-1]$ & DELPHI (LEP2) $\sqrt{s}=189$-$209$ GeV, ${\cal L}=520$ pb$^{-1}$~\cite{Abdallah:2008sf}\\ \hline
    \end{tabular*}
\end{table*}

In this article, we intend to study the $WWZ$ anomalous couplings in $ZW^\pm$ production at the LHC at 
$\sqrt{s}=13$ TeV using the cross sections, forward-backward asymmetries, and 
polarizations asymmetries~\cite{Bourrely:1980mr,Abbiendi:2000ei,Ots:2004hk,Boudjema:2009fz,Aguilar-Saavedra:2015yza,Rahaman:2016pqj,Nakamura:2017ihk}  of $Z$ and $W^\pm$ in the $3l+\cancel{E}_T$ channel.
The polarizations of $Z$ and $W$ have been used  recently for various BSM 
studies~\cite{Renard:2018tae,Renard:2018bsp,Renard:2018lqv,Renard:2018jxe,Renard:2018blr,
Aguilar-Saavedra:2017zkn,Behera:2018ryv} along with studies of anomalous gauge boson 
couplings~\cite{Rahaman:2016pqj,Rahaman:2017qql,Abbiendi:2000ei,Rahaman:2018ujg}. The 
polarizations of $W^\pm/Z$ have been estimated earlier in $ZW^\pm$ production~\cite{Stirling:2012zt,Baglio:2018rcu,Baglio:2019nmc} 
and also have been measured recently at the LHC~\cite{Aaboud:2019gxl} in the SM.
We note that the $ZW^\pm$ processes also contain anomalous couplings other than aTGC, such as the  anomalous $Zq\bar{q}$ couplings, and they affect the measurement of  aTGC~\cite{Zhang:2016zsp,Biekotter:2018rhp,Baglio:2019uty}. However, the main aim of this paper is to demonstrate the usefulness of polarization observables in
probing possible new physics. For simplicity, we
restrict our analysis to possible anomalous couplings  only in the bosonic sector of the SM. 

We will begin in Sect.~\ref{sec:signal-sigma} by providing the estimates of the cross sections
for CMS fiducial phase-space by {\tt MATRIX}~\cite{Grazzini:2017mhc}, \MGvATNLO~\cite{Alwall:2014hca} and investigate their sensitivities  to the anomalous couplings. 
Section~\ref{sec:Pol-Asym} is devoted to polarization asymmetries
of $Z$ and $W$ and the reconstruction of longitudinal momenta of the missing neutrino.
In Sect.~\ref{sec:limits-and-bench}, we perform a simultaneous analysis using the Markov-chain–Monte-Carlo  (MCMC) to obtain
limits on the anomalous couplings along with a toy measurement of {\em non-zero} aTGC and conclude in Sect.~\ref{sec:conclusion}.

%%%%%%%%%%%%%%%%%%%%%%%%%%%%%%%%%%%%%%%%%%%%%%%%%%%%%%%%%%%%%%%%%%%%%%%%%%%%%%%%%%%%%%%%%%%%
\section{Signal cross sections and their sensitivity to anomalous couplings}\label{sec:signal-sigma}
\begin{figure}[h]
    \centering
    \includegraphics[width=1.0\textwidth]{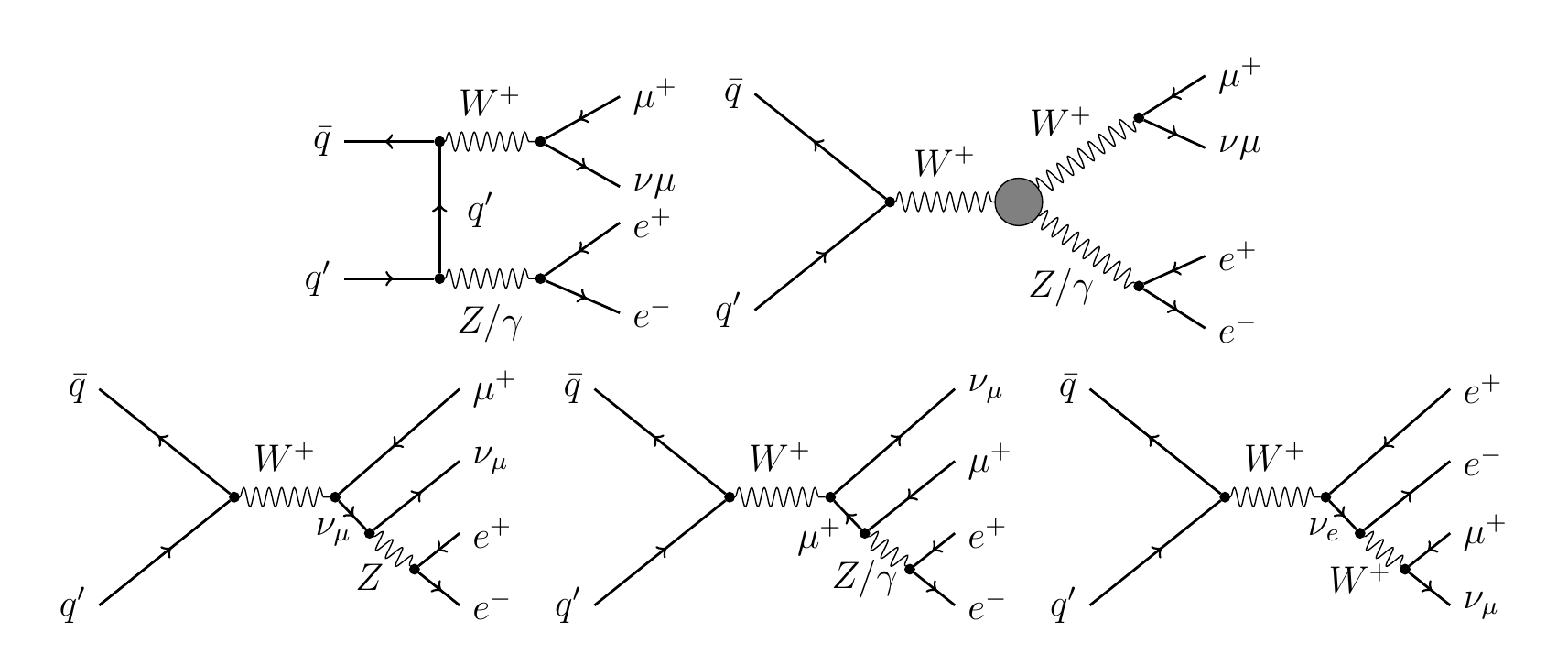}
    \caption{\label{fig:Feynman_WZ_LHC}
Sample of Born level Feynman diagrams for $ZW^+$ production in the $e^+e^-\mu^+\nu_\mu$ channel at the LHC. 
The diagrams for $ZW^-$ can be obtained by charge conjugation. The shaded blob represents the presence of
anomalous $WWV$ couplings on top of SM.}
\end{figure}
\begin{table*}[!ht]\caption{\label{tab:WZ-sigma-SM} The theoretical estimates and experimental measurements of the cross sections of  
        $ZW^\pm$ production in the $e^+e^-\mu^\pm\nu_\mu/\bar{\nu}_\mu$ channels  at $\sqrt{s}=13$ TeV 
        at the LHC for CMS fiducial phase-space. The uncertainties in the  theoretical
        estimates are due to scale variation.}
    \renewcommand{\arraystretch}{1.50}
    \begin{tabular*}{\columnwidth}{@{\extracolsep{\fill}}lllll@{}} \hline
        Process & Obtained at & $\sigma_{\text{LO}}$ (fb)& $\sigma_{\text{NLO}}$ (fb) & $\sigma_{\text{NNLO}}$ (fb) \\ \hline
\multirow{2}{*}{$pp\to e^+e^-\mu^+\nu_\mu$} &{\tt MATRIX} & $22.08_{-6.2\%}^{+5.2\%}$ & $43.95_{-4.3\%}^{+5.4\%}$ & $48.55_{-2.0\%}^{+2.2\%}$  \\ \cline{2-5}
& {\tt mg5\_aMC} & $22.02_{-7.2\%}^{+6.1\%}$ & $43.63_{-6.6\%}^{+6.6\%}$ & ------  \\ \hline
\multirow{2}{*}{$pp\to e^+e^-\mu^-\bar{\nu}_\mu$} & {\tt MATRIX}& $14.45_{-6.7\%}^{+5.6\%}$ & $30.04_{-4.5\%}^{+5.6\%}$ & $33.39_{-2.1\%}^{+2.3\%}$\\ \cline{2-5}
& {\tt mg5\_aMC}& $14.38_{-7.6\%}^{+6.4\%}$ & $29.85_{-6.8\%}^{+6.8\%}$ & ------\\ \hline
$pp \to 3l+\cancel{E}_T $& {\tt MATRIX}~\cite{Grazzini:2017ckn} &$148.4_{-6.4\%}^{+5.4\%}$ &$301.4_{-4.4\%}^{+5.1\%}$ &$334.3_{-2.1\%}^{+2.3\%}$ \\ \hline\hline
        \end{tabular*}
\begin{tabular*}{\columnwidth}{@{\extracolsep{\fill}}llllll@{}}
$pp \to 3l+\cancel{E}_T$&&~~~~~CMS~\cite{Khachatryan:2016tgp} & \hspace{1.2cm}$258.0\pm 8.1\%$ (stat)$^{+7:4\%}_{-7.7\%}$ (syst)$\pm 3.1$ (lumi)& \\ \hline
\end{tabular*}
\end{table*}
\begin{figure}[h!]
	\centering
	\includegraphics[width=0.495\textwidth]{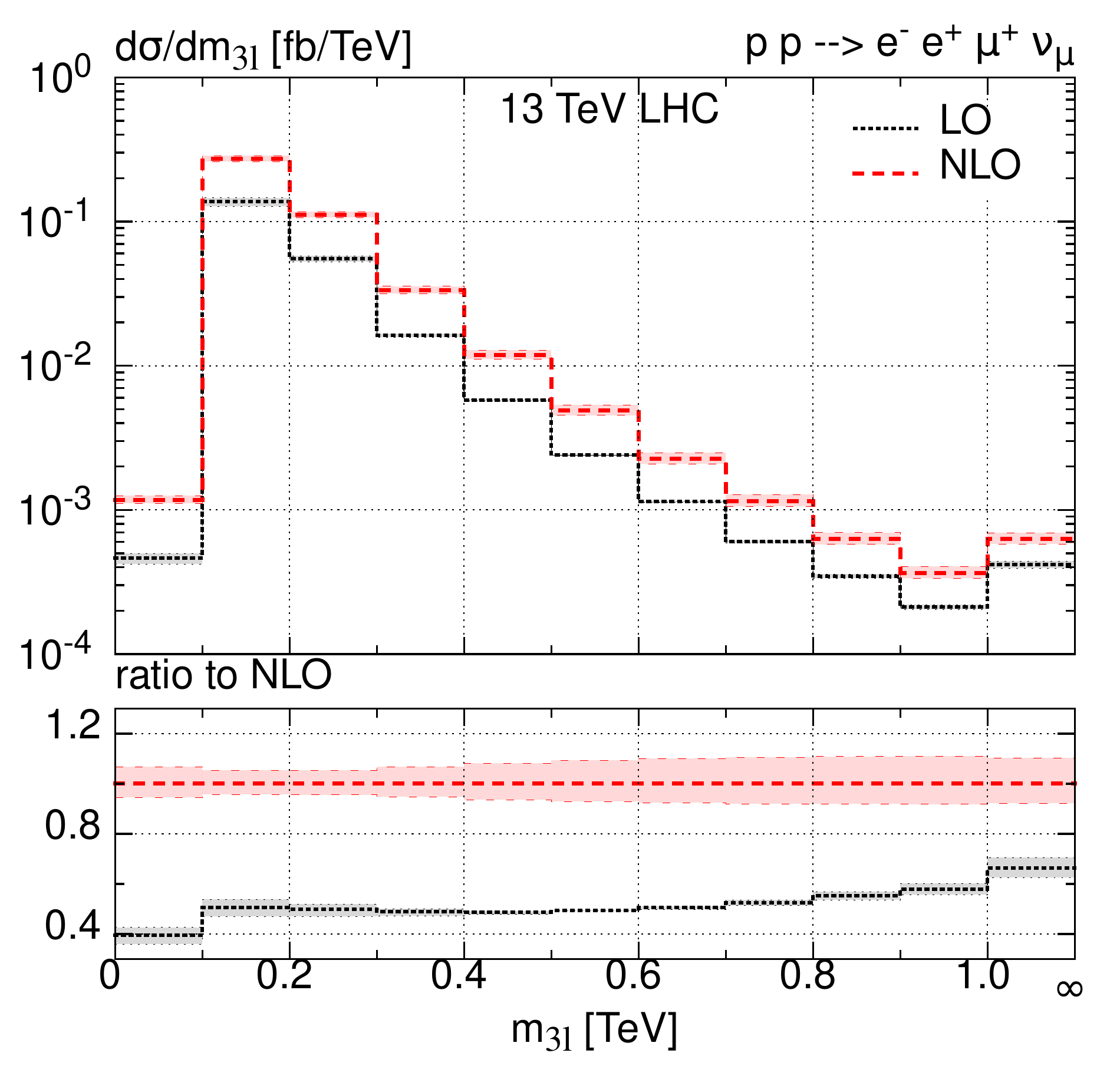}
	\includegraphics[width=0.495\textwidth]{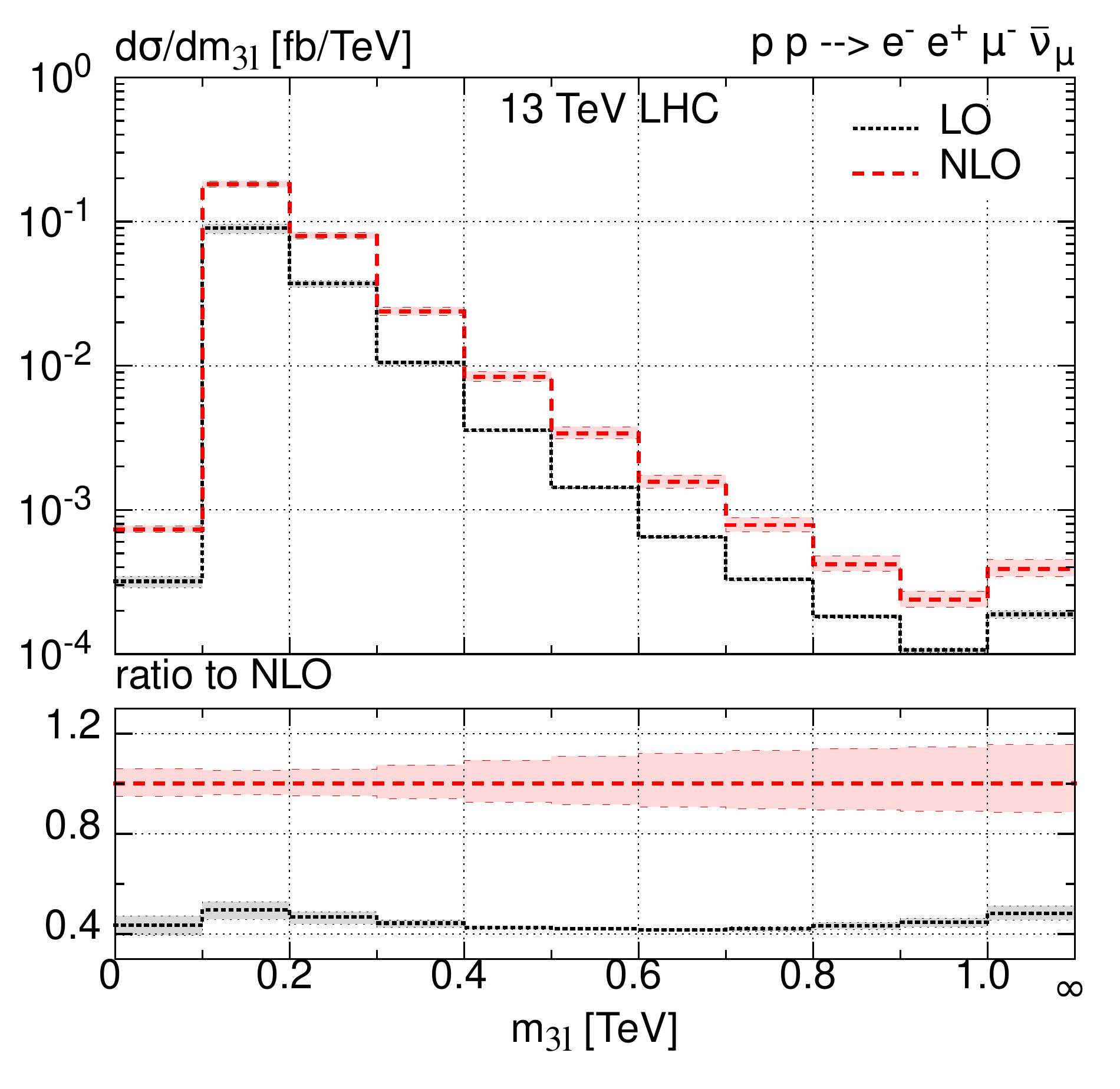}
	\includegraphics[width=0.495\textwidth]{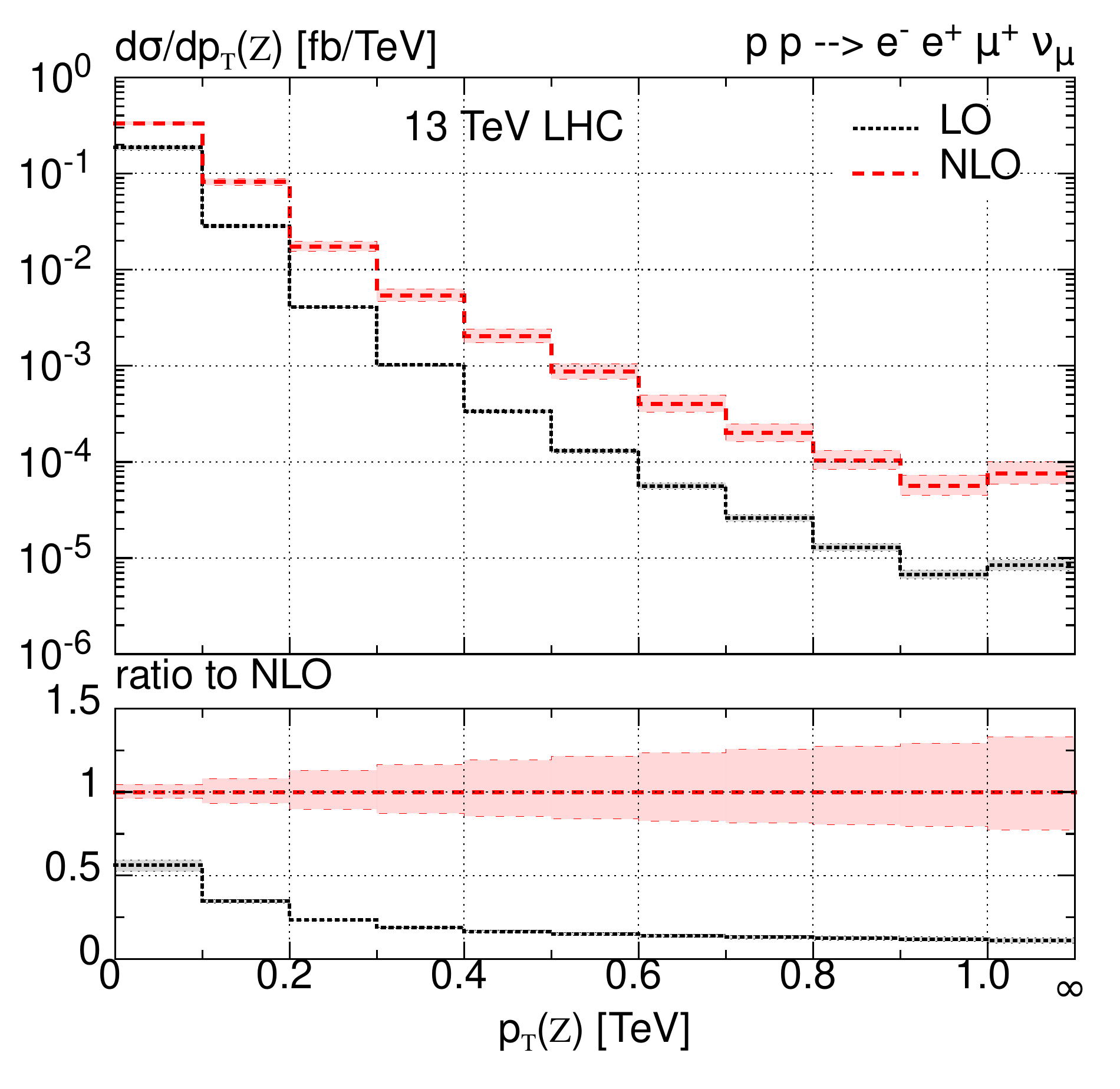}
	\includegraphics[width=0.495\textwidth]{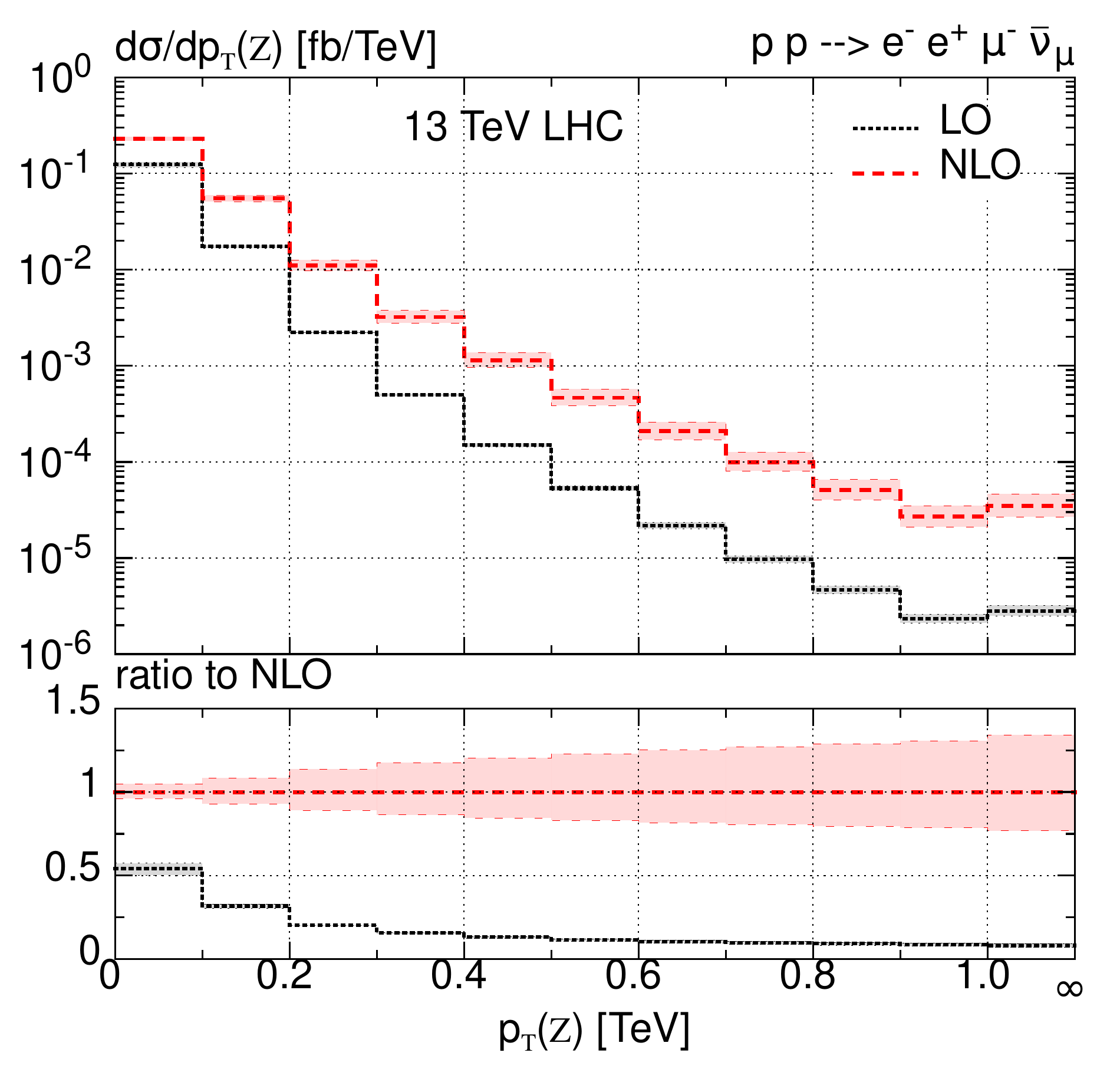}
	\caption{\label{fig:m3l-pTZ-matrx}
		The differential distributions of $m_{3l}$ ({\em top-row}) and $p_T(Z)$ ({\em bottom-row}) in the $ZW^+$ ({\em left-column}) and $ZW^-$ ({\em 
			right-column}) production in the $e^+e^-\mu^\pm+\cancel{E}_T$ channel at the LHC for $\sqrt{s}=13$ TeV at LO and NLO in QCD obtained using {\tt 
			MATRIX}~\cite{Grazzini:2016swo,Grazzini:2017ckn,Grazzini:2017mhc,Cascioli:2011va,Denner:2016kdg,Gehrmann:2015ora,Catani:2012qa,Catani:2007vq}
		for CMS fiducial phase-space.
	}
\end{figure}
The processes of interest are the $ZW^\pm$ production  in the $3l+\cancel{E}_T$ channel
at the LHC. The representative Feynman diagrams at Born level are displayed in Fig.~\ref{fig:Feynman_WZ_LHC}
containing doubly-resonant processes ({\em upper-row}) as well as singly-resonant processes ({\em lower-row}).
The presence of anomalous $WWZ$ couplings is shown by the shaded blob. While this may contain the $WW\gamma$
couplings due to the off-shell $\gamma$, this has been cut out by $Z$ selection cuts, described later.
The leading order result ($\sigma_{\text{LO}}^{\text{th}}=148.4$ fb estimated by {\tt MATRIX} in Ref.~\cite{Grazzini:2017ckn}) 
for the $3l+\cancel{E}_T$ cross section at the LHC is way below the 
measured cross section at the LHC ($\sigma_{\text{exp}}^{\text{CMS}}=258$ fb measured by CMS~\cite{Khachatryan:2016tgp}).
Higher-order corrections   to the tree level result are thus necessary.
The next-to-leading order (NLO) corrections in QCD appear in the vertices connected to the quarks 
(see, Fig.~\ref{fig:Feynman_WZ_LHC}) with either QCD loops or 
QCD radiations from the quarks.
The SM cross sections of $ZW^\pm$ production  in the $e^+e^-\mu^\pm$ channel  obtained
by {\tt MATRIX} and \MGvATNLO~v2.6.4 ({\tt mg5\_aMC}) for $\sqrt{s}=13$ TeV 
for the CMS fiducial phase-phase region are  presented in  Table~\ref{tab:WZ-sigma-SM}.
The CMS fiducial phase-phase region~\cite{Khachatryan:2016tgp} is given by
\begin{eqnarray}\label{eq:CMS_fudicial_region}
p_T(l_{Z,1})>20~\text{GeV},~~p_T(l_{Z,2})>10~\text{GeV},~~p_T(l_{W})>20~\text{GeV} \ \ ,\nonumber\\
|\eta_l|<2.5,~~60~\text{GeV}<m_{l_Z^+l_Z^-}<120~\text{GeV},~~m_{l^+l^-}>4~\text{GeV} \ \ .
\end{eqnarray}
We use the values of the SM input parameters the same as used  in Ref.~\cite{Grazzini:2017ckn} 
(default in {\tt MATRIX}). A fixed renormalization ($\mu_R$) and factorization ($\mu_F$) 
scale of $\mu_R=\mu_F=\mu_0=\frac{1}{2}\left (m_Z+m_W \right)$ is used, and the uncertainties are estimated
by varying the $\mu_R$ and $\mu_F$ in the range of $0.5\mu_0 \le \mu_R , \mu_F \le 2\mu_0$, with the constraint $0.5\le\mu_R/\mu_F\le2$ and shown in Table~\ref{tab:WZ-sigma-SM}. 
We use the NNPDF3.0~\cite{Ball:2014uwa} sets of parton distribution functions (PDFs) with $\alpha_s(m_Z)=0.118$ for our calculations.
The combined result for all leptonic channels given in Ref.~\cite{Grazzini:2017ckn} 
and the  measured cross section by CMS~\cite{Khachatryan:2016tgp} are also presented  in  the same table.
The uncertainties  in the  theoretical estimates are due to scale variation.
The result obtained by {\tt MATRIX} and {\tt mg5\_aMC} matches quite well at both LO and NLO level.
The NLO corrections have increased the LO cross section by up to $100~\%$ and the next-to-next-to-leading order (NNLO) cross section is further
increased by $10~\%$ from the NLO value. 
%It is thus necessary  to include QCD corrections to leading order results.
The higher order corrections to the cross section vary with kinematical variable like $m_{3l}$ and $p_T(Z)$, as shown
in Fig.~\ref{fig:m3l-pTZ-matrx}  obtained by  {\tt MATRIX}~\cite{Grazzini:2016swo,Grazzini:2017ckn,Grazzini:2017mhc,Cascioli:2011va,Denner:2016kdg,Gehrmann:2015ora,Catani:2012qa,Catani:2007vq}. 
 The lower panels display the respective bin-by-bin ratios to the NLO central predictions. 
 The NLO to LO ratio does not appear to be constant over the range of 
 $m_{3l}$ and $p_T(Z)$. Thus a simple  $k$-factor with LO events can not be used 
 as a proxy for NLO events. 
We use results from {\tt mg5\_aMC}, including   NLO  QCD  corrections, for our analysis in the rest of the paper.

\begin{figure}[h!]
    \centering
    \includegraphics[width=0.497\textwidth]{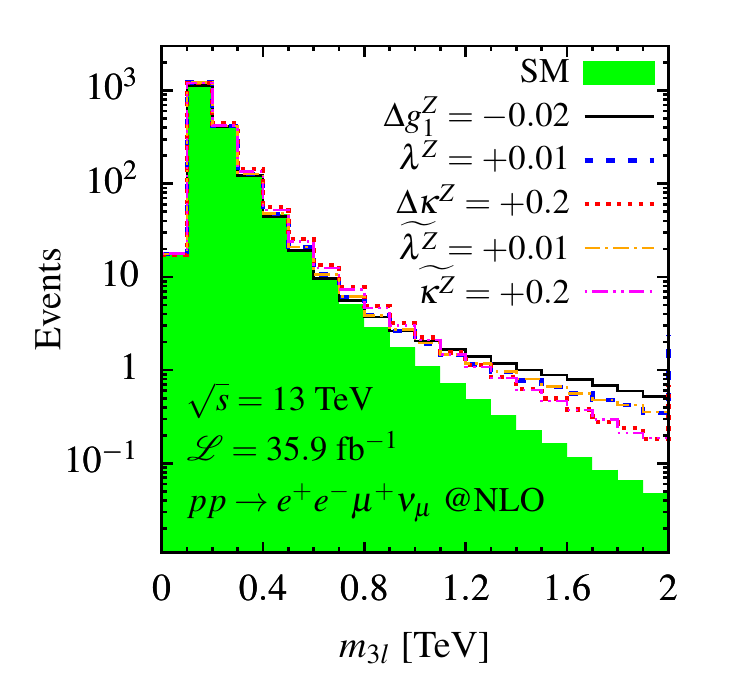}
    \includegraphics[width=0.497\textwidth]{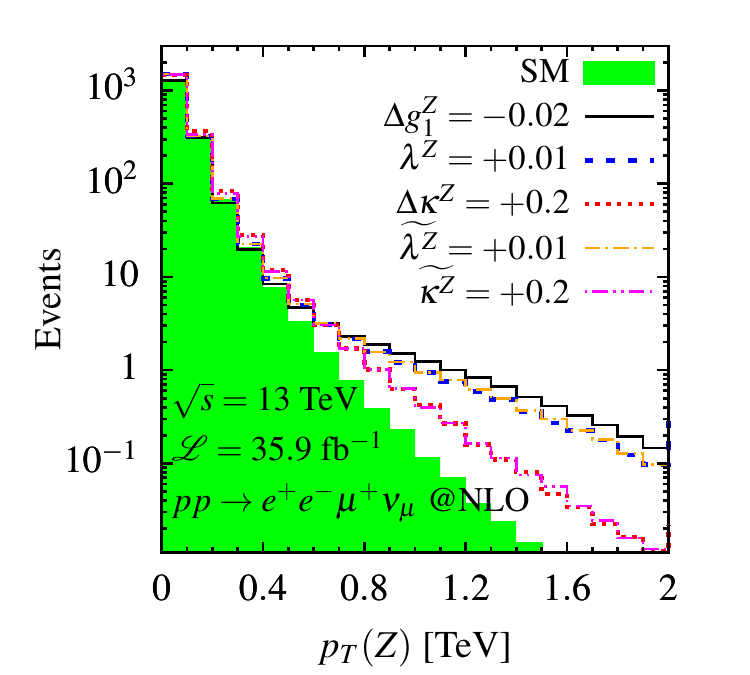}
    \caption{\label{fig:m3l-pTZ-ma5}
       The differential distributions of $m_{3l}$ and $p_T(Z)$ in the $W^+ Z$ production in the $e^+e^-\mu^+\nu_\mu$ channel
        at the LHC at $\sqrt{s}=13$ TeV and ${\cal L}=35.9$ fb$^{-1}$  at NLO in QCD for SM and five benchmark anomalous couplings.
    }
\end{figure}
\begin{figure}[h!]
    \centering
    \includegraphics[width=0.495\textwidth]{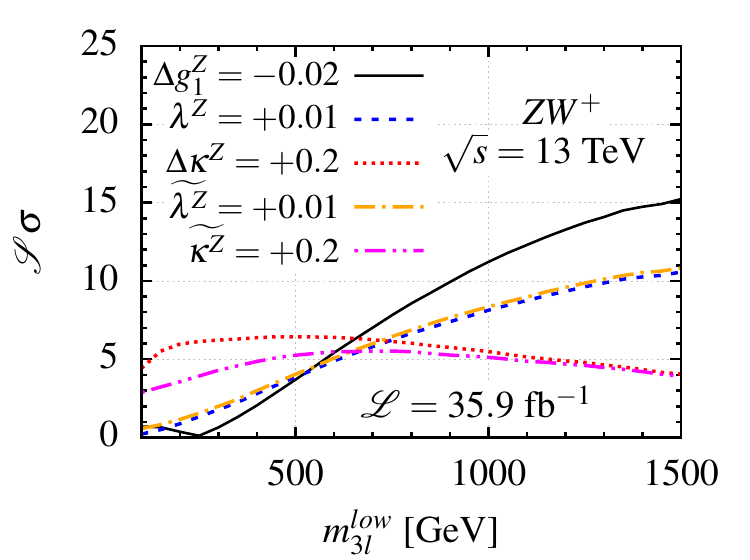}
    \includegraphics[width=0.495\textwidth]{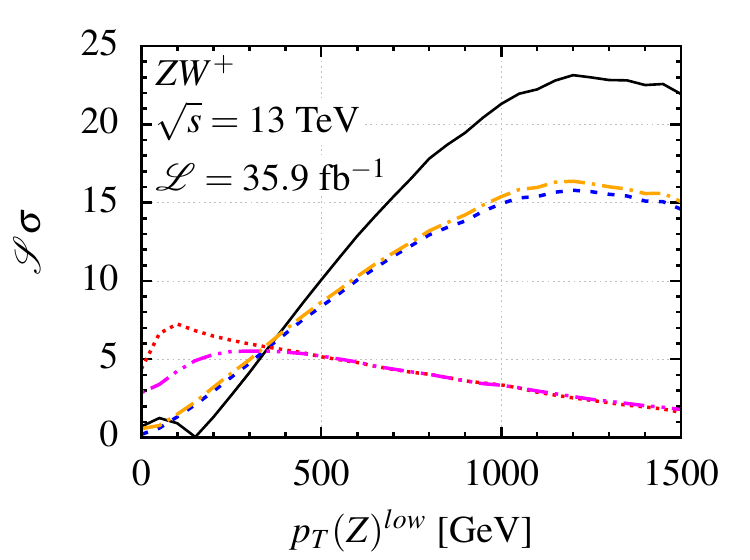}
    \includegraphics[width=0.495\textwidth]{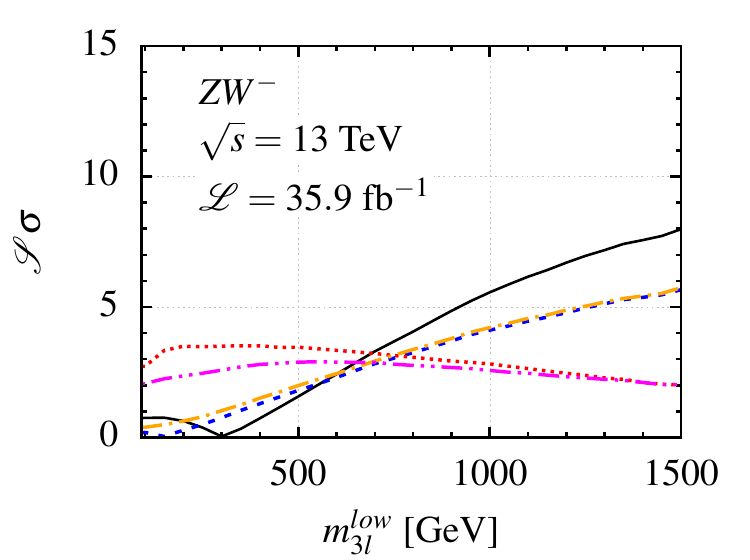}
    \includegraphics[width=0.495\textwidth]{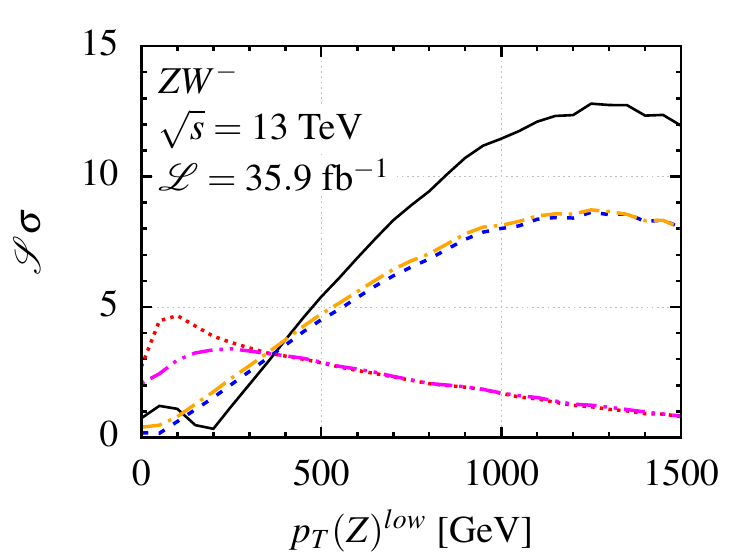}
    \caption{\label{fig:sen-sigma-bench}
        The sensitivities of cross sections  to the five benchmark aTGC as a function of 
        the lower cut on $m_{3l}$ and  $p_T(Z)$ in the $ZW^\pm$ production at the LHC at $\sqrt{s}=13$ TeV and ${\cal L}=35.9$ fb$^{-1}$.
    }
\end{figure}
The signals for the $e^+e^-\mu^+$ and $e^+e^-\mu^-$  are generated  
separately using {\tt mg5\_aMC}  at NLO in QCD for SM as well as  SM including aTGC.
We use the {\tt FeynRules}~\cite{Alloul:2013bka} to generate QCD NLO UFO model of the Lagrangian in Eq.~(\ref{eq:Lagrangian}) 
for {\tt mg5\_aMC}. These 
signals are then used as a proxy for the $3l+\cancel{E}_T$ final state up to a factor of four for the four channels. 
For these, the $p_T$  cut for $e^\pm$ and $\mu^\pm$ are  kept at the same value, i.e., $p_T(l)>10$ GeV.  We use a threshold for the trilepton invariant mass ($m_{3l}$) of $100$ GeV to select the doubly resonant contribution of trilepton final state. Later we will see that a cut of $m_{3l}\ge 100$ GeV  is required  to improve the sensitivities of the observables to the anomalous couplings. 
The event selection cuts for this analysis are thus,
\begin{equation}\label{eq:selection-cuts}
p_T(l)>10~\text{GeV},~~|\eta_l|<2.5,~~60~\text{GeV}<m_{l_Z^+l_Z^-}<120~\text{GeV},~~m_{l^+l^-}>4~\text{GeV},~~m_{3l}>100~\text{GeV} \ \ .
\end{equation}

 We explore the  effect of aTGC in the distributions  of $m_{3l}$ and 
$p_{T}(Z)$ in both $ZW^+$ and $ZW^-$ production and show them in Fig.~\ref{fig:m3l-pTZ-ma5}.
The  distribution of  $m_{3l}$ in the {\em left-panel} and $p_{T}(Z)$ in the 
{\em right-panel} in the $e^+e^-\mu^+\nu_\mu$ channel are shown  for SM ({\it filled}/green) and five anomalous benchmark
couplings\footnote{For each of these benchmark couplings, only one of the couplings is set to non-zero value such that it leads to  $\sim 1 \sigma$ deviation in the total cross section.  More benchmark scenarios ($\sim 100$) with more than one parameters set to  non-zero values at a time are also considered in later sections.}  of $\Delta g_1^Z=-0.02$ ({\it solid}/black), $\lambda^Z=+0.01$ ({\it dashed}/blue), $\Delta\kappa^Z=+0.2$ ({\it dotted}/red),
$\wtil{\lambda^Z}=+0.01$ ({\it dash-dotted}/orange)  and $\wtil{\kappa^Z}=+0.2$ ({\it dashed-dotdotted}/magenta)  
with events normalized to an integrated luminosity of ${\cal L}=35.9$ fb$^{-1}$.  The higher $m_{3l}$ 
and higher $p_T(Z)$ seem to have higher sensitivity to the anomalous couplings
which is due to higher momentum transfer at higher energies, for example see Ref.~\cite{Rahaman:2019mnz}. 
We study the sensitivity of total cross section to the  anomalous 
couplings by varying lower cut on $m_{3l}$  and  $p_T(Z)$  for the above mentioned five benchmark scenarios. The sensitivity of an observable
${\cal O}(c_i)$ to coupling $c_i$ is defined as
\begin{align}\label{eq:def-sensitivity}
{\cal S}{\cal O}(c_i)=\dfrac{|{\cal O}(c_i)-{\cal O}(c_i=0)|}{\delta {\cal O}} \ \ ,
\end{align}
where $\delta {\cal O}$ is the estimated error in ${\cal O}$. For cross sections and  asymmetries, the errors are
\begin{align}
\delta\sigma=\sqrt{\dfrac{\sigma}{{\cal L}} + (\epsilon_\sigma\sigma)^2}~~~~\text{and}~~~
\delta A_i=\sqrt{\dfrac{1-A_i^2}{{\cal L}\times \sigma}+\epsilon_A^2} \ \ ,
\end{align}
where ${\cal L}$ is the integrated luminosity and $\epsilon_\sigma$ and $\epsilon_A$ are the systematic
uncertainties for the cross section and the asymmetries, respectively. 
The sensitivities of the cross sections, ignoring the systematic uncertainty, for the five benchmark cases  (as used in Fig.~\ref{fig:m3l-pTZ-ma5}) are shown in Fig.~\ref{fig:sen-sigma-bench}
 for $ZW^+$ in the {\em upper-row} and for   $ZW^-$ in the {\em lower-row} as a function
 of lower cut  of $m_{3l}$ ({\em left-column}) and  $p_T(Z)$  ({\em right-column}) for luminosity of ${\cal L}=35.9$ fb$^{-1}$. 
 It is clear that the sensitivities increase as the cut  increases  for both 
$m_{3l}$ and  $p_T(Z)$ for couplings $\Delta g_1^Z$, $\lambda^Z$ and $\wtil{\lambda^Z}$,
 while they  decrease just after $\sim 150$ GeV of cuts for the couplings $\Delta\kappa^Z$ and $\wtil{\kappa^Z}$. This can also be seen in Fig.~\ref{fig:m3l-pTZ-ma5} where
 $\Delta\kappa^Z$ and $\wtil{\kappa^Z}$ contribute more than other three couplings for $m_{3l}<0.8$ TeV and $p_T(Z)<0.6$ TeV.
Taking hints from Fig.~\ref{fig:sen-sigma-bench}, we identify four bins in $m_{3l}$-$p_T(Z)$ plane
to maximize the sensitivity of all the couplings. These four bins are given by,
\begin{eqnarray}\label{eq:sigma-twobin}
Bin_{11} &:& 400~\text{GeV}<m_{3l}<1500~\text{GeV},~200~\text{GeV}<p_T(Z)<1200~\text{GeV} \ \ ,\nonumber\\
Bin_{12} &:& 400~\text{GeV}<m_{3l}<1500~\text{GeV},~p_T(Z)>1200~\text{GeV} \ \ ,\nonumber\\
Bin_{21} &:& m_{3l}>1500~\text{GeV},~200~\text{GeV}<p_T(Z)<1200~\text{GeV} \ \ ,\nonumber\\
Bin_{22} &:& m_{3l}>1500~\text{GeV},~p_T(Z)>1200~\text{GeV} \ \ .
\end{eqnarray}
The sensitivities of the cross sections to the benchmark anomalous couplings  are 
 calculated in the said four bins for luminosity of ${\cal L}=35.9$ 
 fb$^{-1}$ and they are shown  in Table~\ref{tab:sen-sigma-bench-twobin} in both 
$ZW^+$ and $ZW^-$ productions. 
As expected, we see that  $Bin_{22}$ has the higher sensitivity to couplings $\Delta g_1^Z$, $\lambda^Z$ and $\wtil{\lambda^Z}$, 
while $Bin_{11}$ has higher, but comparable sensitivity to couplings $\Delta\kappa^Z$ 
and $\wtil{\kappa^Z}$. 
The simultaneous cuts on both the variable have 
increased the sensitivity by a significant amount
as compared to the individual cuts. For example,  Fig.~\ref{fig:sen-sigma-bench} 
shows that cross section in $ZW^+$ 
has a maximum sensitivity of $15$ and $22$  on $\Delta g_1^Z = -0.02$  for individual $m_{3l}$  
and  $p_T(Z)$ lower cuts, respectively.  While imposing  simultaneous lower cuts on both the
variable,  the same sensitivity increases to $44.5$ (in $Bin_{22}$).
\begin{table}\caption{\label{tab:sen-sigma-bench-twobin} The sensitivities of the cross sections  on the five benchmark aTGC in 
the four bins (see Eq.~(\ref{eq:sigma-twobin})) of $m_{3l}$ and  $p_T(Z)$ in the $ZW^\pm$ 
productions at the LHC at $\sqrt{s}=13$ TeV and ${\cal L}=35.9$ fb$^{-1}$. }
\renewcommand{\arraystretch}{1.50}
\begin{tabular*}{\textwidth}{@{\extracolsep{\fill}}|c|cccc|cccc|@{}}\hline
& \multicolumn{4}{c|}{$ZW^+$}& \multicolumn{4}{c|}{$ZW^-$} \\ \hline
aTGC                           & $Bin_{11}$&$Bin_{12}$ &$Bin_{21}$&$Bin_{22}$ & $Bin_{11}$&$Bin_{12}$ &$Bin_{21}$ & $Bin_{22}$ \\ \hline
$\Delta g_1^Z = -0.02$         &$1.17 $&$1.14 $&$7.52 $ &$44.5 $ & $0.32 $ &$2.10 $&$3.95 $ & $23.19 $   \\ \hline
$\lambda^Z = 0.01$             &$3.08 $&$5.37 $&$6.08 $ &$26.2 $ & $ 1.58$ &$2.63 $&$3.32 $ & $13.68 $ \\ \hline
$\Delta\kappa^Z = 0.2$         &$8.52 $&$0.50 $&$3.28 $ &$4.87 $ & $5.01 $ &$0.15 $&$1.64 $ & $2.40 $ \\ \hline
$\widetilde{\lambda^Z} = 0.01$ &$3.20 $&$5.56 $&$6.18 $ &$27.2 $ & $1.70 $ &$2.69 $&$3.37 $ & $13.83 $ \\ \hline
$\widetilde{\kappa^Z} = 0.2$   &$6.50 $&$0.60 $&$3.15 $ &$4.89 $ & $3.86 $ &$0.22 $&$1.65 $ & $2.36 $ \\ \hline
\end{tabular*}
\end{table} 

At the LHC, the other contributions to the   $3l+\cancel{E}_T$ channel  come from the production of  $ZZ$, $Z\gamma$, $Z+j$, $t\bar{t}$, $Wt$, $WW+j$, 
$t\bar{t}+V$, $tZ$, $VVV$ as has been studied by CMS~\cite{Khachatryan:2016tgp,Sirunyan:2019bez} 
and ATLAS~\cite{Aaboud:2016yus,Aaboud:2019gxl}.
The total non-$ZW$ contributions listed above is about $40~\%$ of the $ZW$ contributions~\cite{Khachatryan:2016tgp}.
We include these extra contributions to the cross sections while estimating limits on the anomalous couplings in Sect.~\ref{sec:limits-and-bench}.

%%%%%%%%%%%%%%%%%%%%%%%%%%%%%%%%%%%%%%%%%%%%%%%%%%%%%%%%%%%%%%%%%%%%%%%%%%%%%%%%%%%%%%%%%%%%%%%%%%%
\section{Polarization observables of $Z$ and $W^\pm$ along with other angular asymmetries}\label{sec:Pol-Asym}
Being a spin-$1$ particle, the $Z/W$ ($V$) offers eight additional observables related to their
eight degrees of polarizations apart from their production cross sections. 
 The angular distributions of the daughter particle
reveal the polarizations of the mother particle $V$. 
The normalized  decay angular distribution of the daughter fermion $f$ ($l_Z/l_W$) from the 
decay of $V$ is given by~\cite{Boudjema:2009fz}
\begin{eqnarray} \label{eq:angular_distribution_ZW}
\frac{1}{\sigma} \ \frac{d\sigma}{d\Omega_f} &=&\frac{3}{8\pi} \left[
\left(\frac{2}{3}-(1-3\delta) \ \frac{T_{zz}}{\sqrt{6}}\right) + \alpha \ p_z
\cos\theta_f %\right.\nonumber\\
+ \sqrt{\frac{3}{2}}(1-3\delta) \ T_{zz} \cos^2\theta_f
\right.\nonumber\\
&+&\left(\alpha \ p_x + 2\sqrt{\frac{2}{3}} (1-3\delta)
\ T_{xz} \cos\theta_f\right) \sin\theta_f \ \cos\phi_f \nonumber\\
&+&\left(\alpha \ p_y + 2\sqrt{\frac{2}{3}} (1-3\delta)
\ T_{yz} \cos\theta_f\right) \sin\theta_f \ \sin\phi_f \nonumber\\
&+&(1-3\delta) \left(\frac{T_{xx}-T_{yy}}{\sqrt{6}} \right) \sin^2\theta_f
\cos(2\phi_f)\nonumber\\
&+&\left. \sqrt{\frac{2}{3}}(1-3\delta) \ T_{xy} \ \sin^2\theta_f \
\sin(2\phi_f) \right].
\end{eqnarray}
Here $\theta_f$, $\phi_f$ are the polar and the azimuthal orientation of the  fermion $f$,
in the rest frame of the particle ($V$) with its would be momentum along the  $z$-direction.
For massless final state fermions, we have  $\delta=0$ and $\alpha=(R_f^2- L_f^2)/ (R_f^2+L_f^2)$
for $Z$ with $Zf\bar{f}$ coupling to be $\gamma^\mu \left(L_f \ P_L + R_f \ P_R \right)$
and $\alpha=-1$ for $W^\pm$. The quantities $p_x$, $p_y$, and $p_z$ are the three vector  polarizations
 and $T_{xy}$, $T_{xz}$, $T_{yz}$, $T_{xx}-T_{yy}$, and $T_{zz }$ are the five independent 
 tensor polarizations of the particle $V$.
These  polarizations $p_i$ and $T_{ij}$ are calculable from  asymmetries constructed from the 
decay angular information of lepton using Eq.~(\ref{eq:angular_distribution_ZW}). 
For example, the polarization parameters $p_z$ and
$T_{xz}$ can be calculated from the asymmetries $A_z$ and  $A_{xz}$, respectively, as 
\begin{eqnarray}\label{eq:Asym_Az_Axxyy}
A_z&=&\frac{1}{\sigma}\left[
\int _{0}^{\frac{\pi}{2}}  
\dfrac{d\sigma}{d\theta_f}d\theta_f
-\int _{\frac{\pi}{2}}^{\pi}
\dfrac{d\sigma}{d\theta_f}
d\theta_f \right]
\equiv  \dfrac{\sigma(\cos\theta_f>0)-\sigma(\cos\theta_f<0)}{\sigma(\cos\theta_f>0)+\sigma(\cos\theta_f<0)}\nonumber\\
&=&\frac{3 \alpha  p_z}{4} \ \ ,\nonumber\\
A_{xz}&=&\frac{1}{\sigma}\l[\l(\int _{\theta =0}^{\frac{\pi}{2} }\int _{\phi =-\frac{\pi}{2}}^{\frac{\pi}{2}} \dfrac{d\sigma}{d\Omega_f}d\Omega_f +
\int _{\theta=\frac{\pi}{2}}^{\pi}\int _{\phi=\frac{\pi}{2}}^{\frac{3\pi}{2}}
\dfrac{d\sigma}{d\Omega_f}d\Omega_f\r) \r.\nonumber\\
&&\l.- 
\l(\int _{\theta =0}^{\frac{\pi}{2} }\int _{\phi=\frac{\pi}{2}}^{\frac{3\pi}{2}}
\dfrac{d\sigma}{d\Omega_f}d\Omega_f+
\int _{\theta=\frac{\pi}{2}}^{\pi}\int _{\phi=-\frac{\pi}{2}}^{\frac{\pi}{2}}
\dfrac{d\sigma}{d\Omega_f}d\Omega_f
   \r)
\r]\nonumber\\
&\equiv&  \dfrac{\sigma(\cos\theta_f\cos\phi_f>0)-\sigma(\cos\theta_f\cos\phi_f<0)}{\sigma(\cos\theta_f\cos\phi_f>0)+\sigma(\cos\theta_f\cos\phi_f<0)} \nonumber \\
&=&\frac{2}{\pi } \sqrt{\frac{2}{3}} (1-3 \delta ) T_{xz} \ \ .
\end{eqnarray}
Similarly one can construct asymmetries corresponding to each of the other polarizations
$p_i$ and $T_{ij}$; see Ref.~\cite{Rahaman:2016pqj} for details. 

The $Z$ and the $W^\pm$ bosons produced in the $ZW^\pm$ production are not forward-backward
symmetric owing to only a $t$-channel diagram and not having an $u$-channel diagram (see Fig.~\ref{fig:Feynman_WZ_LHC}). 
These provide an extra observable,  the forward-backward asymmetry defined as
\begin{equation}\label{eq:def-Afb}
A_{fb}^V=\dfrac{\sigma(\cos\theta_V>0)-\sigma(\cos\theta_V<0)}{\sigma(\cos\theta_V>0)+\sigma(\cos\theta_V<0)} \ \ ,
\end{equation}
$\theta_V$ is the production angle of the $V$ w.r.t. the colliding quark-direction. 
\begin{figure}
    \centering
    \includegraphics[width=0.496\textwidth]{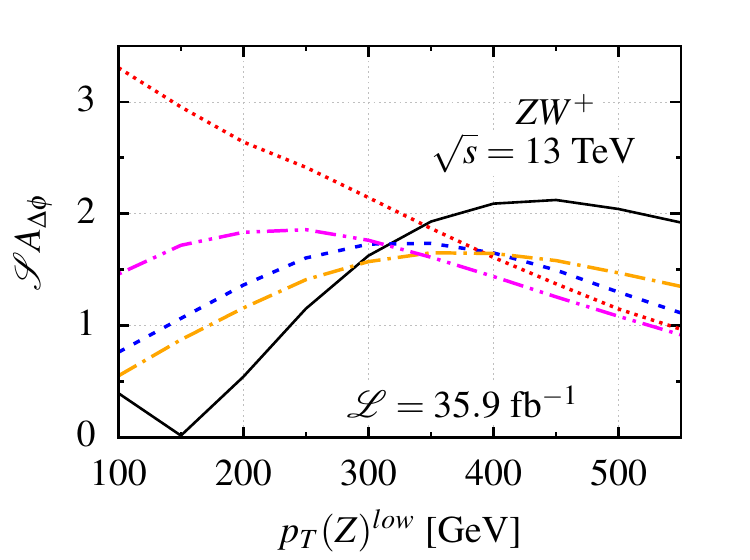}
    \includegraphics[width=0.496\textwidth]{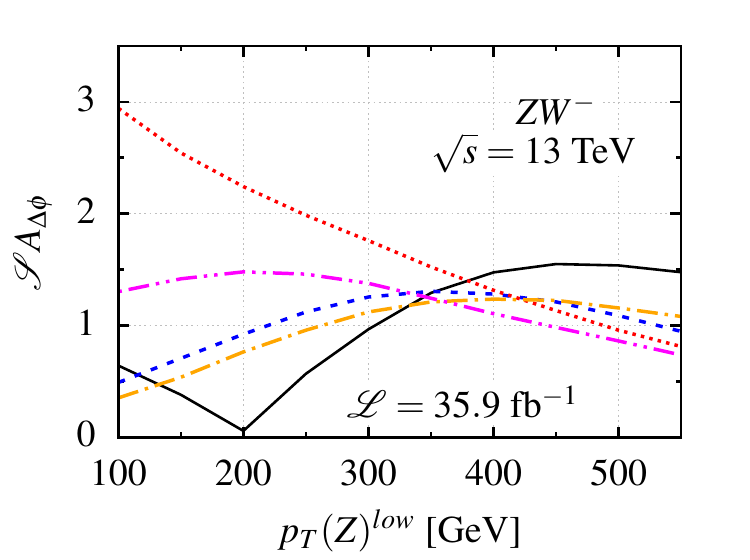}
    \caption{\label{fig:dist_deltaphi}
The sensitivity of the asymmetry $A_{\Delta\phi}$  on the five benchmark aTGC as a function of 
the lower cut on   $p_T(Z)$ in the $ZW^\pm$ production at the LHC at $\sqrt{s}=13$ TeV and ${\cal L}=35.9$ fb$^{-1}$.
The legend labels are same as in Fig.~\ref{fig:sen-sigma-bench}.}
\end{figure}
One more angular variable
sensitive to aTGC is the angular separation of the lepton $l_W$ from $W^\pm$ and the $Z$ in the
transverse plane, i.e, 
\begin{equation}\label{eq:deltaphi_lw_Z}
\Delta\phi(l_W,Z) = \cos^{-1}\l(\dfrac{\vec{p}_T(l_W).\vec{p}_T(Z)}{p_T(l_W) p_T(Z)}\r) \ \ .
\end{equation}
One can construct an asymmetry based on the $\Delta\phi$  as,
\begin{equation}\label{eq:deltaphi_Asym}
A_{\Delta\phi}= \dfrac{\sigma\l(\cos\l( \Delta\phi(l_W,Z) \r)>0\r)-\sigma\l(\cos\l( \Delta\phi(l_W,Z) \r)<0\r)}{\sigma\l(\cos\l( \Delta\phi(l_W,Z) \r)>0\r)+\sigma\l(\cos\l( \Delta\phi(l_W,Z) \r)<0\r)} \ \ .
\end{equation}
The sensitivities of $A_{\Delta\phi}$ to the five benchmark 
aTGC are shown in Fig.~\ref{fig:dist_deltaphi} as a function of lower cuts on $p_T(Z)$ in both 
$ZW^\pm$ for luminosity of ${\cal L}=35.9$ fb$^{-1}$.  A choice of $p_T(Z)^{low}=300$ GeV appears to
be an optimal choice for sensitivity for all the couplings. The $m_{3l}$ cut, however, reduces the sensitivities
to all the aTGC.

To construct the asymmetries, we need to set a  reference frame and assign the leptons to the correct mother
spin-$1$ particle. For the present process with missing neutrino, we face a set of challenges in constructing the asymmetries. These are
discussed below.
\paragraph{Selecting $Z$ candidate leptons}
The $Z$ boson momentum
is required to be reconstructed to obtain all the asymmetries which require the right pairing of the $Z$
boson leptons $l_Z^+$ and  $l_Z^-$.
Although the opposite flavor channels $e^+e^-\mu^\pm/\mu^+\mu^-e^\pm$ are safe, the same 
flavor channels $e^+e^-e^\pm/\mu^+\mu^-\mu^\pm$ suffer ambiguity to select the right $Z$ boson
candidate leptons. The right paring of
leptons for the $Z$ boson in the same flavored channel is possible with $\ge 96.5~\%$ accuracy for
$m_{3l}>100$ GeV and  $\ge 99.8~\%$ accuracy for $m_{3l}>550$ GeV in
both SM and benchmark aTGC by requiring a smaller value of $|m_Z-m_{l^+l^-}|$. This small  miss pairing is 
neglected to  use the $2e\mu\nu_\mu$ channel as a proxy 
for a $3l+\cancel{E}_T$ final state with good enough accuracy.

\paragraph{The  reconstruction of neutrino momentum}
The other major issue is to obtain the asymmetries related to $W^\pm$ bosons, which require to 
reconstruct their momenta. As the neutrino from $W^\pm$ goes missing, reconstruction
of $W^\pm$ boson momenta is possible with a two-fold ambiguity using the transverse missing 
energy $\cancel{p_T}/\cancel{E_T}$ and the on-shell $W$ mass ($m_W$) constrain. The two solutions for
the longitudinal momentum of the missing neutrino are given by
\begin{equation}\label{eq:pznu-solution}
{p_z(\nu)}_\pm = \frac{-\beta p_z(l_W)\pm E(l_W)\sqrt[]{D}}{p_T^2(l_W)} 
\end{equation}
with 
\begin{equation}\label{eq:pznu-sol-Dbeta}
D = \beta^2 -p_T^2(\nu)p_T^2(l_W)~,~~~
\beta=m_W^2 + p_x(l_W) p_x(\nu) + p_y(l_W) p_y(\nu) \ \ .
\end{equation}
\begin{figure}
	\centering
	\includegraphics[width=0.496\textwidth]{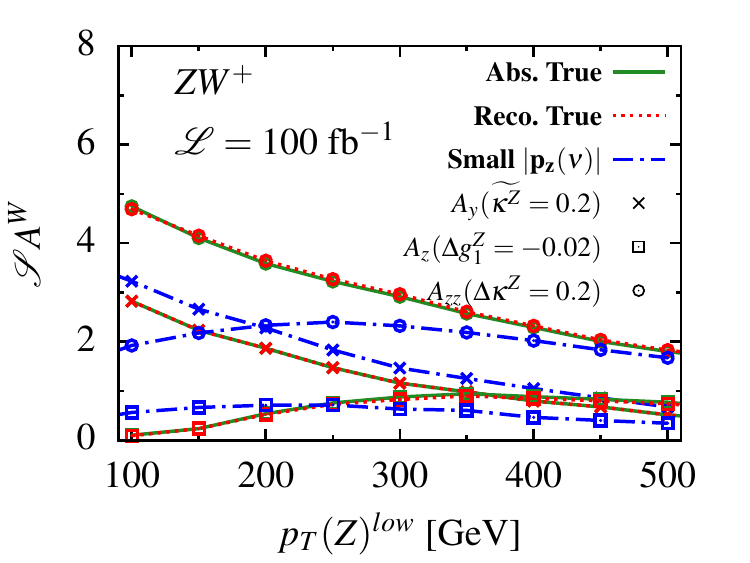}
	\includegraphics[width=0.496\textwidth]{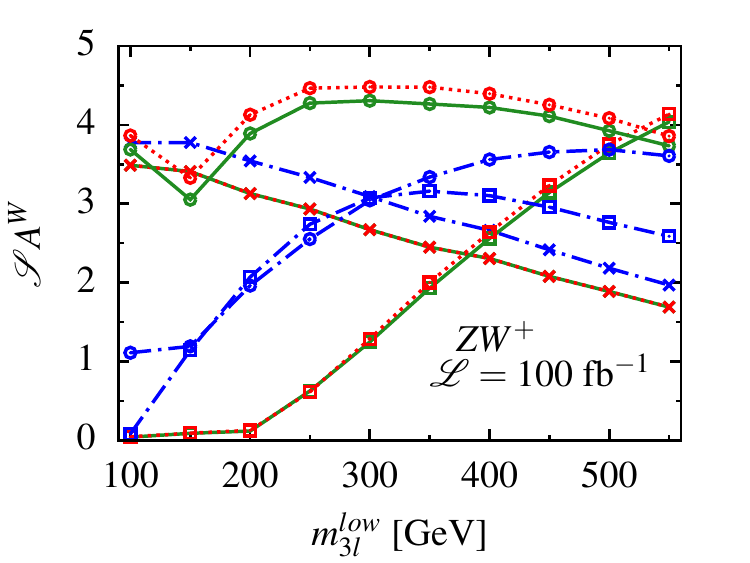}
	\includegraphics[width=0.496\textwidth]{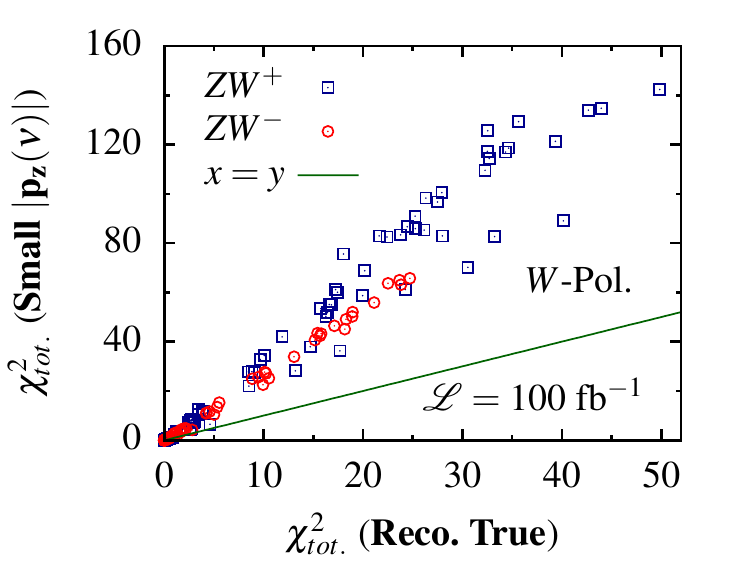}
	\caption{\label{fig:Sen-Wpol-Bench-TrueVsReco}
 The sensitivity of some polarization asymmetries of $W^+$ ($ZW^+$)  
on some benchmark aTGC for three scenarios: with absolute truth (\textbf{Abs.~True}) information of 
neutrino in {\em solid}/blue lines, with the close to true reconstructed solution of neutrino 
(\textbf{Reco.~True}) in {\em dotted}/red lines and with the smaller $|p_z(\nu)|$ to be the true 
solution (\textbf{Small}~$\mathbf{|p_z(\nu)|}$) in {\em dash-dotted}/blue lines as a function of 
the lower cut on $p_T(Z)$ ({\em top-left-panel}) and $m_{3l}$ ({\em top-right-panel}) at $\sqrt{s}=13$ TeV and ${\cal L}=100$ fb$^{-1}$. The scatter plot of  the total $\chi^2$ for  about $100$ aTGC points
using all the asymmetries of $W^\pm$ for \textbf{Reco.~True} in $x$-axis with \textbf{Small}~$\mathbf{|p_z(\nu)|}$  in $y$-axis is shown in the {\em bottom-panel}.
}
\end{figure}
Because the $W$ is not produced on-shell all the time, among the two solutions of longitudinal neutrino momenta, one of them will be closer to the true value, and another will be  far from the true value.
There are no suitable selector or discriminator to select the correct solution from 
the two solutions. Even if we substitute the Monte-Carlo truth $m_W$ to 
solve for $p_z(\nu)$, we don't have any discriminator to distinguish between the two solutions $p_z(\nu)_\pm$. 
The smaller value of $|p_z(\nu)|$ corresponds to the correct solution only for $\approx 65\%$ times on average
in $ZW^+$ and little lower in $ZW^-$ production. One more discriminator, which 
is  $||\beta_Z|-|\beta_W||$, the smaller value of this can choose the correct solution a little over
the boundary, i.e., $\approx 55\%$. We have tried machine-learning approaches (artificial neural network) to select the 
correct solutions, but the accuracy was not better than $ 65\%$. 
In some cases, we have $D<0$ with the on-shell $W$. For these cases, either one can throw those events (which affects the distribution and statistics), or one can 
vary the $m_W$ from its central value to have $D>0$. Here, we follow the latter. 
 So, as the best available option, we choose the smaller value of  $|p_z(\nu)|$
to be the correct solution to reconstruct the $W$ boson momenta.
 At this point,
it becomes important to explore the effect of reconstruction on asymmetries and their
sensitivities to aTGC. To this end, we consider three scenarios: 
\begin{description}
	\item[\textbf{Abs.~True}] The first thing is to use the  Monte-Carlo truth events and estimate the asymmetries in the lab frame.
The  observables in this scenario are directly related to the dynamics up to a rotation of frame~\cite{Bourrely:1980mr,V.:2016wba,Velusamy:2018ksp}.

	\item[\textbf{Reco.~True}] Using the pole mass of $W$ in Eq.~(\ref{eq:pznu-sol-Dbeta}) and choosing the solution closer to
	the Monte-Carlo true value is the best that one can do in  reconstruction. 
		The goal of any reconstruction algorithm would be to become as close to this scenario as possible. 
		
	\item[\textbf{Small}~$\mathbf{|p_z(\nu)|}$]   This choice is the best available realistic algorithm which we will be using for the analysis.
\end{description}

The values of reconstructed  asymmetries and hence polarizations   get shifted  from  \textbf{Abs.~True} case.
 In the case of \textbf{Reco.~True}, the shifts are roughly constant, while
in the case of \textbf{Small}~$\mathbf{|p_z(\nu)|}$, the shifts are not constant over
varying lower cuts on $m_{3l}$ and $p_T(Z)$ due to the $35~\%$ wrong choice. It is, thus, expected that
the reconstructed sensitivities to aTGC remain the same  in \textbf{Reco.~True} and change in the \textbf{Small}~$\mathbf{|p_z(\nu)|}$ case when compared to the \textbf{Abs.~True} case. In the \textbf{Small}~$\mathbf{|p_z(\nu)|}$ reconstruction case, sensitivities 
of some asymmetries to aTGC are less than that of   the \textbf{Abs.~True} case, while they are higher for some other asymmetries.
This is illustrated
in Fig.~\ref{fig:Sen-Wpol-Bench-TrueVsReco} ({\it top-row}) comparing the sensitivity of some polarization
asymmetries of $W^+$, e.g., $A_y$ to $\wtil{\kappa^Z}=+0.2$ in cross ($\times$) points, $A_z$ to 
$\Delta g_1^Z=-0.02$ in square  ($\boxdot$) points,  and  $A_{zz}$ to  
$\Delta\kappa^Z=+0.2$ in circular ($\odot$) points for the three scenarios of \textbf{Abs.~True} ({\em solid}/blue line), \textbf{Reco.~True} ({\em dotted}/red) and \textbf{Small}~$\mathbf{|p_z(\nu)|}$ ({\em dash-dotted}/blue) for varying lower cuts on $p_T(Z)$ and $m_{3l}$ in $ZW^+$ production with a luminosity of ${\cal L}=100$ 
fb$^{-1}$. The sensitivities are roughly the same for \textbf{Abs.~True} and 
\textbf{Reco.~True} reconstruction in all the asymmetries for both $p_T(Z)$ and $m_{3l}$ cuts.
In the \textbf{Small}~$\mathbf{|p_z(\nu)|}$ reconstruction case, sensitivity is smaller for $A_{zz}$,  higher for $A_y$,  and it depends on cut for $A_z$ when compared to the \textbf{Abs.~True} case.  When all the $W$ asymmetries are combined, the total 
 $\chi^2$  is higher in the     \textbf{Small}~$\mathbf{|p_z(\nu)|}$ case compared to  the \textbf{Reco.~True} case  for about $100$ chosen
 benchmark points; see Fig.~\ref{fig:Sen-Wpol-Bench-TrueVsReco} ({\em bottom-panel}).   
Here, a  total $\chi^2$  of all the  asymmetries of $W$ ($A_i^W$) for a set of benchmark points ($\{c_i\}$) is given by
\begin{equation}
\chi^2(A_i^W)(\{c_i\})=\sum_{j}^{N=9} \l({\cal S}A_j^W(\{c_i\}) \r)^2.
\end{equation}
The said increment of $\chi^2$ is observed in both $W^+Z$ ($\boxdot$/blue) and $W^-Z$ ($\odot$/red)  production processes. 
So even if we are not able to reconstruct the $W$ and hence its polarization observables correctly, realistic effects end up enhancing the overall sensitivity 
of the observables to the aTGC.

\paragraph{Reference $z$-axis for polarizations}
The other challenge to obtain the polarization of $V$ is that one needs a reference axis ($z$-axis) 
to get the momentum direction of $V$, which is not possible at the LHC as it is a
symmetric collider. Thus, for the asymmetries related to $Z$ boson, we consider the direction of total visible longitudinal momenta as an unambiguous choice for
positive $z$-axis. 
For the case of $W$,  the direction of the reconstructed boost is used as a proxy for the positive $z$-axis. The latter choice is inspired by the fact that
in $q^\prime\bar{q}$ fusion the quark is supposed to have larger momentum
than the anti-quark at the LHC, thus the above proxy could  stand statistically for the direction of 
the quark direction.

\paragraph{List of observables}\label{para:list-of-obs}
The set of observables used in this analysis are,
\begin{itemize}
\item[$\sigma_i$]: The cross sections in four bins ($4$),
\item[$A_{pol}^Z$]: Eight polarization asymmetries of $Z$ ($8$), 
\item[$A_{fb}^Z$]: Forward-backward asymmetry of $Z$ ($1$), 
\item[$A_{\Delta\phi}$]: Azimuthal asymmetry ($1$),
\item[$A_{pol}^W$]: Eight polarization asymmetries of reconstructed $W$ ($8$),
\item[$A_{fb}^W$]: Forward-backward asymmetry of reconstructed $W$\footnote{We note that the forward-backward asymmetry of $Z$ and $W$ are ideally the same in the CM frame. However, since we measure the $Z$ and $W$ $\cos\theta$  w.r.t. different quantity, i.e., visible $p_z$ for $Z$ and reconstructed boost for $W$, they are practically different and we use them as two independent observables.} ($1$),
\end{itemize}
which make a total of $N({\cal O})=(4+8+1+1+8+1)\times 2=46$ observables including both processes.
All the asymmetry from $Z$ side and all the asymmetries from $W$ side are termed as $A_i^Z$ and 
$A_i^W$, respectively, for the latter uses.
The total $\chi^2$ for all observables would be the quadratic sum of 
sensitivities (Eq.~(\ref{eq:def-sensitivity}))  given by
\begin{equation}\label{eq:tot-chi2}
\chi^2_{tot}(c_i) = \sum_{j}^{N=46} \l({\cal S}{\cal O}_j(c_i) \r)^2 \ \ .
\end{equation}
We use these set of observables in some chosen kinematical region to obtain limits on aTGC in the next section.

%%%%%%%%%%%%%%%%%%%%%%%%%%%%%%%%%%%%%%%%%%%%%%%%%%%%%%%%%%%%%%%%%%%%%%%%%%%%%%%%%%%%%%%%%%%%%
\section{Measurement of the anomalous couplings}\label{sec:limits-and-bench}
We study the sensitivity of all the ($N({\cal O})=46$) observables   for varying lower cuts on $m_{3l}$ and $p_T(Z)$
separately as well as simultaneously (grid scan in step of $50$ GeV in each direction) for the chosen benchmark anomalous couplings.
The maximum sensitivities are observed for simultaneous lower cuts on $m_{3l}$ and $p_T(Z)$
given in  Table~\ref{tab:m3lpTZcut-on-Asym} for all the asymmetries in both $ZW^\pm$ processes. Some of these
cuts can be realised from Fig.~\ref{fig:dist_deltaphi} \& \ref{fig:Sen-Wpol-Bench-TrueVsReco}.
The SM values of the asymmetries of $Z$ and $W$ and their corresponding polarizations
for the selection  cuts ({\tt sel.cut} in Eq.~(\ref{eq:selection-cuts})) and for the optimized  cuts ({\tt opt.cut} in Table~\ref{tab:m3lpTZcut-on-Asym}) are listed in Table~\ref{tab:SM-values-Asym-Pol}
in appendix~\ref{app:SM-values-Asym} for completeness. 
We use the cross sections in the four bins and  all the asymmetries with the optimized cuts to obtain limits on the anomalous 
couplings for both effective vertices and  effective operators. We  use the semi-analytical expressions for the observables fitted with the simulated data from {\tt mg5\_aMC}. The details of the fitting procedures  are described in appendix~\ref{app:fitting}.
The uncertainty on the cross sections and asymmetries are taken as $\epsilon_\sigma=20~\%$ and $\epsilon_A=2~\%$, respectively consistent with the analysis by CMS~\cite{Khachatryan:2016tgp} and ATLAS~\cite{Aaboud:2019gxl}. We note that these uncertainties are not considered
in the previous sections for qualitative analysis and optimization of cuts.
\begin{table}[hb!]\caption{\label{tab:m3lpTZcut-on-Asym}  The list of  optimized lower cuts ({\tt opt.cut}) in GeV on  ($m_{3l}$,$p_T(Z)$)  for various asymmetries to maximize their sensitivity to the anomalous couplings.}
	\renewcommand{\arraystretch}{1.50}
	\begin{tabular*}{\textwidth}{@{\extracolsep{\fill}}|cccc|@{}}\hline
%	\begin{tabular*}{|c|c|c|c|}\hline
		${\cal O}$      & $Z$ in $ZW^+$ & $Z$ in $ZW^-$ & $W^\pm$ in $ZW^\pm$  \\ \hline
		$A_x$           & $(200,100)$   & $(100,150)$   &  $(250,0)$         \\ 
		$A_y$           & $(150,100)$   & $(100,100)$   &  ''                  \\
		$A_z$           & $(550,50)$    & $(100,250)$   &  ''                  \\
		$A_{xy}$        & $(150,100)$   & $(150,100)$   &  ''                  \\
		$A_{xz}$        & $(150,0)$     & $(200,50)$    &  ''                  \\
		$A_{yz}$        & $(100,50)$    & $(100,0)$     &  ''                  \\
		$A_{x^2-y^2}$   & $(400,150)$   & $(300,100)$   &  ''                  \\
		$A_{zz}$        & $(550,0)$     & $(300,400)$   &  ''                  \\
		$A_{fb}$        & $(300,0)$     & $(550,0)$     &  ''                \\ \hline
	\end{tabular*}            
	\begin{tabular*}{\textwidth}{@{\extracolsep{\fill}}|ccc|@{}}\hline
		& $ZW^+$       &   $ZW^-$                            \\\hline
		$A_{\Delta\phi}$& $(100,300)$   & $(100,300)$                      \\ \hline
	\end{tabular*}
	%    \end{scriptsize}
\end{table}

\begin{figure}
	\centering
	\includegraphics[width=0.496\textwidth]{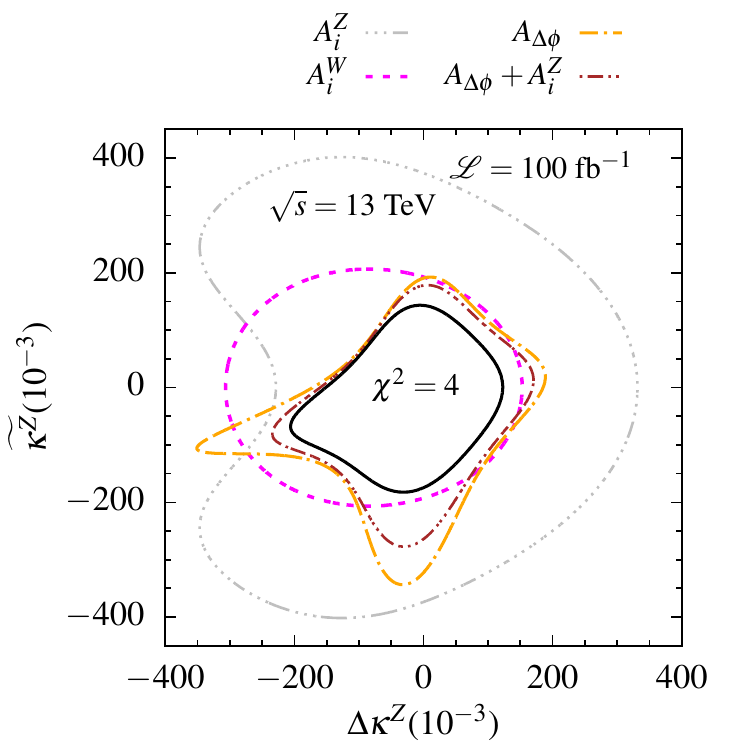}
	\includegraphics[width=0.496\textwidth]{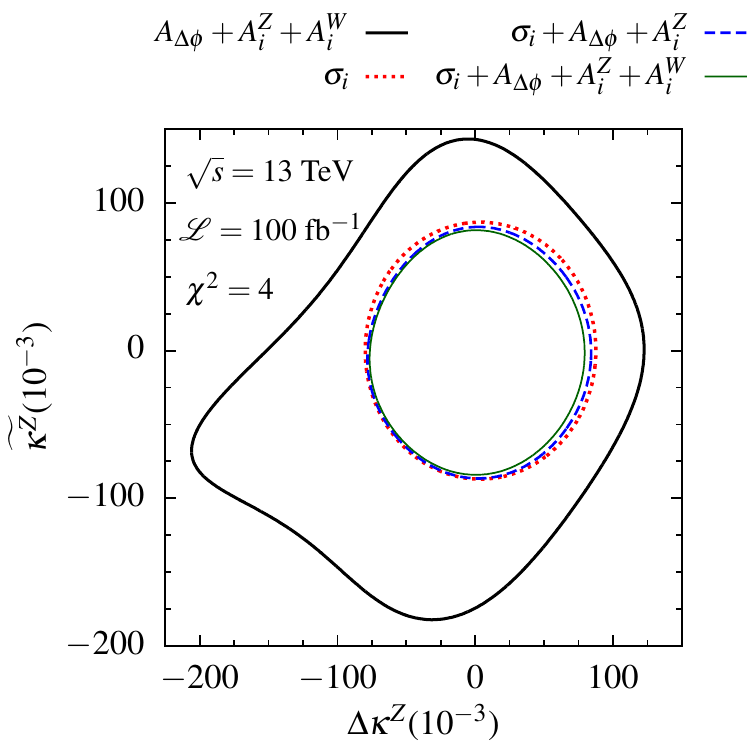}
	\caption{\label{fig:chi2-sigZW-contour-Lum100fb}
		The $\chi^2=4$ contours are shown in the $\Delta\kappa^Z$--$\wtil{\kappa^Z}$ plane with different asymmetries and their combinations in the {\em left-panel},  various combinations of the  cross sections  and asymmetries in the {\em right-panel} 
		for $\sqrt{s}=13$ TeV and  ${\cal L}=100$ fb$^{-1}$. The contour for $A_{\Delta\phi}+A_i^Z+A_i^W$ ({\em thick-solid}/black line ) is repeated in both {\em panel} for comparison.
	}
\end{figure}

The sensitivities of all the observables to the aTGC are studied by varying one-parameter, two-parameter, and all-parameter at a time in the optimized cut region.
We look at the $\chi^2=4$ contours in the $\Delta\kappa^Z$-$\wtil{\kappa^Z}$ plane  for a luminosity of ${\cal L}=100$ fb$^{-1}$ for various combinations
of asymmetries and cross sections and show them in Fig.~\ref{fig:chi2-sigZW-contour-Lum100fb}.
We observe that the $Z$-asymmetries ($A_i^Z$) are weaker than the $W$-asymmetries ($A_i^W$);   $A_i^W$ provides
very symmetric limits, while $A_i^Z$ has a sense of directionality.
The  $A_{\Delta\phi}$ is better than both $A_i^Z$ and $A_i^W$ in most of the directions in $\Delta\kappa^Z$-$\wtil{\kappa^Z}$ plane.
After combining  $A_i^Z$,   $A_i^W$ and $A_{\Delta\phi}$, we get a tighter contours; but the shape is dictated by $A_{\Delta\phi}$.
We see (Fig.~\ref{fig:chi2-sigZW-contour-Lum100fb} {\em right-panel})  that the cross sections have  higher sensitivity compared to
the asymmetries to the aTGC.  
The cross sections dominate constraining  the couplings, while the contribution from  the asymmetries  remain sub-dominant at best. 
Although the directional constraints provided by the asymmetries  get washed away when combined with the cross sections, they are expected to remain prominent to extract {\em non-zero} couplings   should a deviation from the SM  be observed. This possibility is discussed in subsection~\ref{subsec:rol-of-asym}.

\subsection{Limits on the couplings}
\begin{table}\caption{\label{tab:simul-limits-Lag-Op-OpLag} The list of simultaneous limits from the MCMC
		at $95~\%$ BCI on  the effective vertex couplings $c_i^{\cal L}$  and the effective operator 
		couplings $c_i^{\cal O}$  along with translated limits on effective vertices $c_i^{{\cal L}_g}$ for
		various luminosities with the notation $_{ lower~limit}^{ higher~limit}\equiv [ lower~limit,  higher~limit]$.}
	\renewcommand{\arraystretch}{1.50}
	\begin{tabular*}{\textwidth}{@{\extracolsep{\fill}}cccccc@{}}\hline
		$c_i^{\cal L}$ $(10^{-3})$                  &$35.9$ fb$^{-1}$       & $100$ fb$^{-1}$        & $300$ fb$^{-1}$        & $1000$ fb$^{-1}$ & $3000$ fb$^{-1}$      \\ \hline
		$\Delta g_1^Z$                  &$_{  -4.20 }^{+   2.15 }$ &$_{   -3.47 }^{+   1.50 }$ &$_{   -2.92 }^{+   0.963}$ &$_{   -2.48 }^{+  0.565}$ & $_{-2.17}^{+0.318}$ \\\hline
		$\lambda^Z $                    &$_{  -2.24 }^{+   2.11 }$ &$_{   -1.78 }^{+   1.66 }$ &$_{   -1.42 }^{+   1.30 }$ &$_{   -1.14 }^{+   1.01}$& $_{-0.931}^{+0.811}$ \\\hline
		$\Delta\kappa^Z$                &$_{ -83.0 }^{+  83.5   }$ &$_{  -64.1  }^{+  66.6  }$ &$_{  -47.9  }^{+  52.8  }$ &$_{  -34.2  }^{+  42.1 }$& $_{-27.2}^{+36.0}$ \\\hline
		$\widetilde{\lambda^Z}$         &$_{  -2.19 }^{+   2.19 }$ &$_{   -1.74 }^{+   1.72 }$ &$_{   -1.38 }^{+   1.36 }$ &$_{   -1.09 }^{+   1.09}$& $_{-0.884}^{+0.883}$ \\\hline
		$\widetilde{\kappa^Z}$          &$_{ -88.4 }^{+  86.2   }$ &$_{  -70.4  }^{+  67.5  }$ &$_{  -54.9  }^{+  51.8  }$ &$_{  -43.2  }^{+  40.1 }$& $_{-36.7}^{+33.9}$ \\\hline
		$c_i^{\cal O}$  (TeV$^{-2}$) &&&                                                                                                                  \\\hline
		$\frac{c_{WWW}}{\Lambda^2}$               &$_{ -0.565 }^{+  0.540 }$ &$_{  -0.445 }^{+  0.426 }$ &$_{  -0.365 }^{+  0.327 }$ &$_{  -0.258 }^{+  0.257}$& $_{-0.238}^{+0.200}$ \\\hline
		$\frac{c_{W}}{\Lambda^2}$                 &$_{ -0.747 }^{+  0.504 }$ &$_{  -0.683 }^{+  0.397 }$ &$_{  -0.624 }^{+  0.274 }$ &$_{  -0.390 }^{+  0.196}$& $_{-0.381}^{+0.138}$ \\\hline
		$\frac{c_{B}}{\Lambda^2}$                 &$_{-67.1   }^{+ 67.8   }$ &$_{ -59.2   }^{+ 60.1   }$ &$_{ -52.6   }^{+ 47.6   }$ &$_{ -33.3   }^{+ 30.9  }$& $_{-30.1}^{+27.0}$ \\\hline
		$\frac{c_{\wtil{WWW}}}{\Lambda^2}$        &$_{ -0.514 }^{+  0.516 }$ &$_{  -0.430 }^{+  0.415 }$ &$_{  -0.342 }^{+  0.339 }$ &$_{  -0.244 }^{+  0.252}$& $_{-0.216}^{+0.209}$ \\\hline
		$\frac{c_{\wtil{W}}}{\Lambda^2}$          &$_{-68.5   }^{+ 69.2   }$ &$_{ -60.4   }^{+ 61.2   }$ &$_{ -52.0   }^{+ 52.7   }$ &$_{ -32.7   }^{+ 34.2  }$& $_{-30.0}^{+31.0}$ \\\hline
		$c_i^{{\cal L}_g}$ $(10^{-3})$ &&&                                                                                                          \\\hline
		$\Delta g_1^Z$                  &$_{  -3.10 }^{+   2.10 }$ &$_{   -2.84 }^{+   1.65 }$ &$_{   -2.59 }^{+   1.14 }$ &$_{   -1.62 }^{+  0.814}$& $_{-1.58}^{+0.576}$ \\\hline
		$\lambda^Z $                    &$_{  -2.31 }^{+   2.21 }$ &$_{   -1.82 }^{+   1.74 }$ &$_{   -1.49 }^{+   1.34 }$ &$_{   -1.06 }^{+   1.05}$& $_{-0.975}^{+0.822}$ \\\hline
		$\Delta\kappa^Z$                &$_{ -63.4  }^{+  62.1  }$ &$_{  -56.4  }^{+  54.6  }$ &$_{  -44.8  }^{+  48.3  }$ &$_{  -29.1  }^{+  30.6 }$& $_{-25.6}^{+27.6}$ \\\hline
		$\widetilde{\lambda^Z}$         &$_{  -2.10 }^{+   2.11 }$ &$_{   -1.76 }^{+   1.70 }$ &$_{   -1.40 }^{+   1.39 }$ &$_{   -1.00 }^{+   1.03}$& $_{-0.882}^{+0.857}$ \\\hline
		$\widetilde{\kappa^Z}$          &$_{ -64.5  }^{+  63.8  }$ &$_{  -57.1  }^{+  56.3  }$ &$_{  -49.1  }^{+  48.4  }$ &$_{  -31.9  }^{+  30.5 }$& $_{-28.9}^{+28.0}$ \\\hline
	\end{tabular*} 
\end{table}
\begin{figure}
	\centering
	\includegraphics[width=1.0\textwidth]{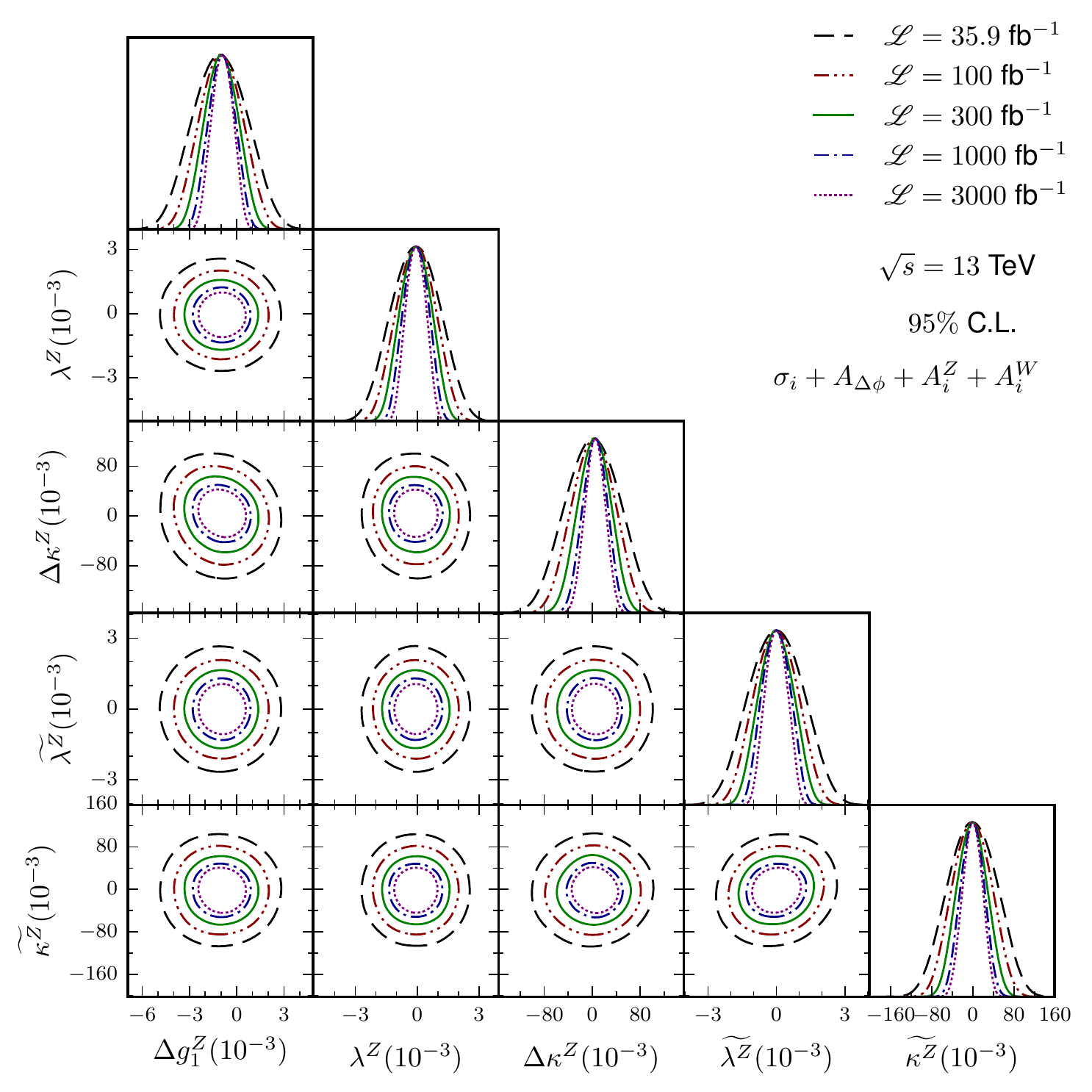}
	\caption{\label{fig:mcmc_Limit_Lag}
	All the  marginalised $1D$ projections and $2D$ projections at $95~\%$ BCI from the MCMC in triangular array
		for the effective vertices ($c_i^{\cal L}$)  for  various luminosities at $\sqrt{s}=13$ TeV using all the observables.}
\end{figure}
\begin{figure}[h!]
	\centering
	\includegraphics[width=1.0\textwidth]{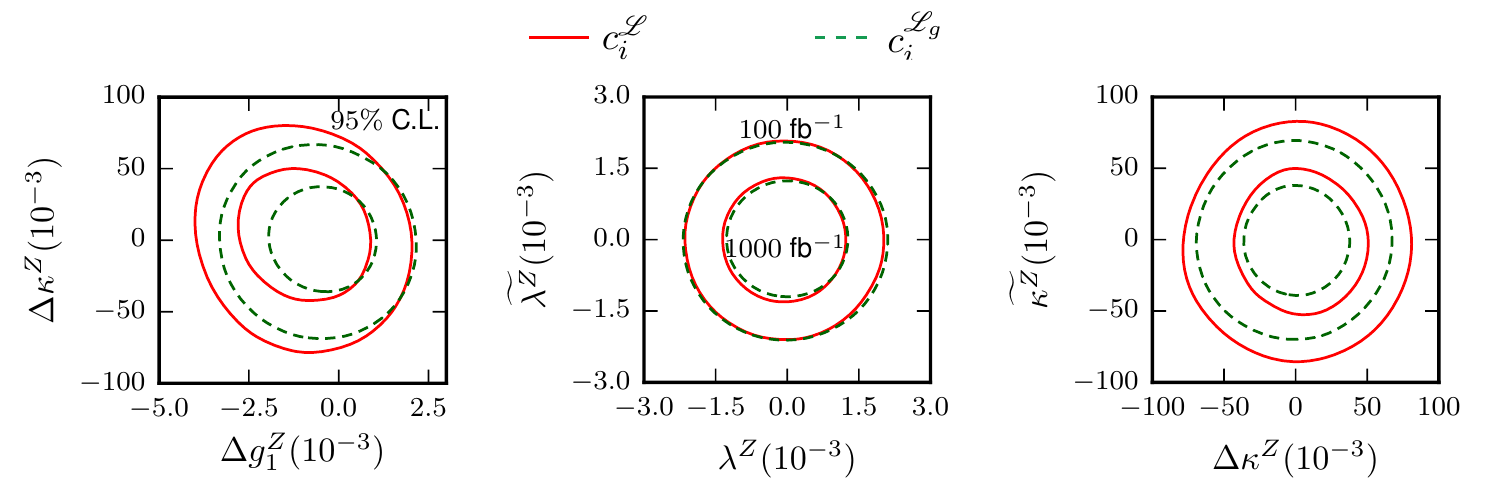}
	\caption{\label{fig:mcmc_Lag_vs_OpLag} The marginalised $2D$ projections at $95~\%$ BCI from the MCMC
		in the $\Delta g_1^Z$-$\Delta\kappa^Z$, $\lambda^Z$-$\wtil{\lambda^Z}$, and 
		$\Delta\kappa^Z$-$\wtil{\kappa^Z}$ planes are shown in {\em solid}/red when the effective vertex factors
		($c_i^{\cal L}$) are treated independent, while shown in {\em dashed}/green
		when the operators are treated independent ($c_i^{{\cal L}_g}$) for luminosities ${\cal L}=1000$ fb$^{-1}$
		(two inner contours) and ${\cal L}=100$ fb$^{-1}$ (two outer contours)
		at $\sqrt{s}=13$ TeV using all the observables.}
\end{figure}

\begin{figure}
	\centering
	\includegraphics[width=0.49\textwidth]{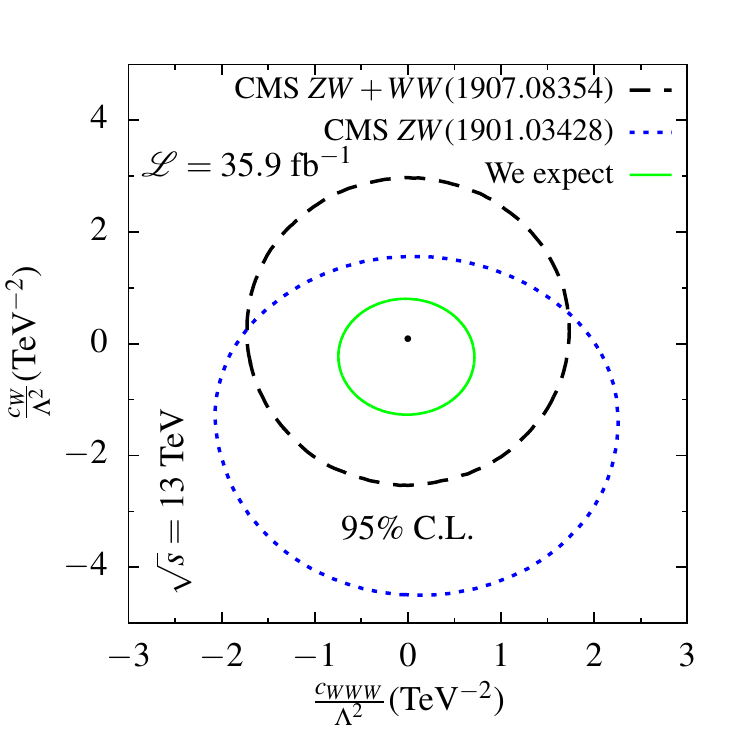}
	\includegraphics[width=0.5\textwidth]{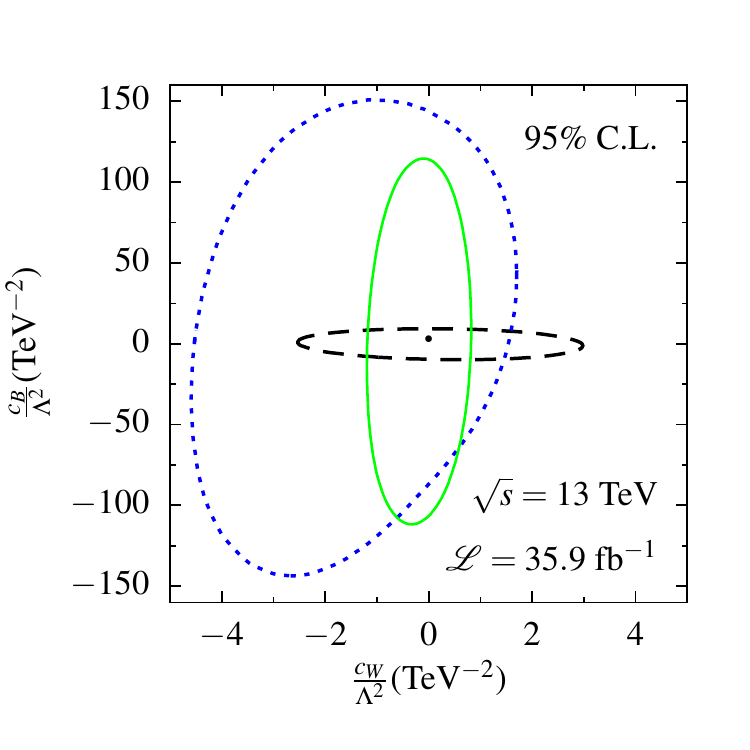}
	\caption{\label{fig:comparision_with_CMS_Op}
		The two parameter $95~\%$ C.L. contours in the $c_{WWW}/\Lambda^2$--$c_W/\Lambda^2$ plane ({\em left-panel}) and $c_{W}/\Lambda^2$--$c_B/\Lambda^2$ plane ({\em right-panel}) for  our estimate (We~expect) in {\em solid}/green lines, for CMS $ZW+WW$ in {\em dashed}/black lines  and for  CMS $ZW$ in {\em dotted}/blue lines
		at $\sqrt{s}=13$ TeV and  ${\cal L}=35.9$ fb$^{-1}$ using all the observables.
	}
\end{figure}
We extract simultaneous limits  on all the anomalous couplings using all the  observables
using the MCMC method. We perform this analysis in two ways:  ($i$) vary effective vertex factors 
couplings ($c_i^{\cal L}$)  and
($ii$) vary effective operators couplings ($c_i^{\cal O}$) and translate them in to effective vertex  factors couplings ($c_i^{{\cal L}_g}$)  using Eq.~(\ref{eq:Operator-to-Lagrangian}).
The   $95~\%$ BCI (Bayesian confidence interval) obtained on aTGC are listed  in Table~\ref{tab:simul-limits-Lag-Op-OpLag}  for five choices of integrated  luminosities:  ${\cal L}=35.9$ fb$^{-1}$, 
${\cal L}=100$ fb$^{-1}$, ${\cal L}=300$ fb$^{-1}$, ${\cal L}=1000$ fb$^{-1}$ and ${\cal L}=3000$ fb$^{-1}$.
The correlation among the parameters are studied  (using {\tt GetDist}~\cite{Antony:GetDist}) and 
they  are shown in Fig.~\ref{fig:mcmc_Limit_Lag} along with $1D$ projections for effective vertex  factors.
The limits on the couplings get tighter as the luminosity is increased, as it should be. The shape of 
the contours are very circular in all two-parameter projections as the cross sections dominate
in constraining the aTGC. The same conclusions are drawn when effective operators are
varied as independent parameters. 
The limits on $c_i^{{\cal L}_g}$ are tighter compared to the limits on $c_i^{\cal L}$ (see Table~\ref{tab:simul-limits-Lag-Op-OpLag});
the comparison between them are shown in the two-parameter marginalised plane  in Fig.~\ref{fig:mcmc_Lag_vs_OpLag}
in $\Delta g_1^Z$-$\kappa^Z$,  $\lambda^Z$-$\wtil{\lambda^Z}$ and $\kappa^Z$-$\wtil{\kappa^Z}$ planes as representative
for luminosity ${\cal L}=100$ fb$^{-1}$ (outer contours) and ${\cal L}=1000$ fb$^{-1}$ (inner contours). 
The limits and the contours are roughly the same in  $\lambda^Z$-$\wtil{\lambda^Z}$  plane.  The contours are more symmetric
around the SM for $c_i^{{\cal L}_g}$ compared to $c_i^{\cal L}$, e.g.; see $\Delta g_1^Z$-$\kappa^Z$ plane.
The  limits obtained here for luminosity $35.9$ fb$^{-1}$ 
are better than the experimentally observed limits at the LHC given in 
Table~\ref{tab:aTGC_constrain_form_collider}  except on $c_B$ and hence 
on $\Delta\kappa^Z$. This is because
the LHC analysis~\cite{Sirunyan:2019gkh} uses $WW$ production on top of $ZW$ production,
whereas we only use $ZW$ production process. But our  limits on the couplings are better when compared
with the $ZW$ production process alone at the LHC~\cite{Sirunyan:2019bez}. 
In Fig.~\ref{fig:comparision_with_CMS_Op}, we present the comparison of limits obtained by the CMS analyses with 
$ZW+WW$~\cite{Sirunyan:2019gkh} process and $ZW$~\cite{Sirunyan:2019bez}
with our estimate with two parameter $95~\%$ BCI contours  in the $c_{WWW}/\Lambda^2$--$c_W/\Lambda^2$ plane ({\em left-panel}) and $c_{W}/\Lambda^2$--$c_B/\Lambda^2$ plane ({\em right-panel}). The contour in the plane $c_{WWW}/\Lambda^2$--$c_W/\Lambda^2$
in our estimate (We~expect) ({\em solid}/green line) is tighter compared to both  
CMS $ZW+WW$ ({\em dashed}/black line) and CMS $ZW$ analyses ({\em dotted}/blue line). 
This is because we use binned cross sections  in the analysis.
The limit on the couplings $c_B/\Lambda^2$ ({\em right-panel}) on the other hand  
is tighter, yet comparable, with CMS $ZW$ and weaker than the CMS $ZW+WW$ analysis because
the $ZW$ process itself is less sensitive to $c_W$.

%%%%%%%%%%%%%%%%%%%%%%%%%%%%%%%%%%%%%%%%%%%%%%%%%%%%%%%%%%%%%%%%%%%%%%%%%%%%%%%%%%%%%%%%%%%%%%%%%%%%%%%%%%%%%%%
\subsection{The role of asymmetries in parameter extraction}\label{subsec:rol-of-asym}
\begin{figure}[h]
	\centering
	\includegraphics[width=1\textwidth]{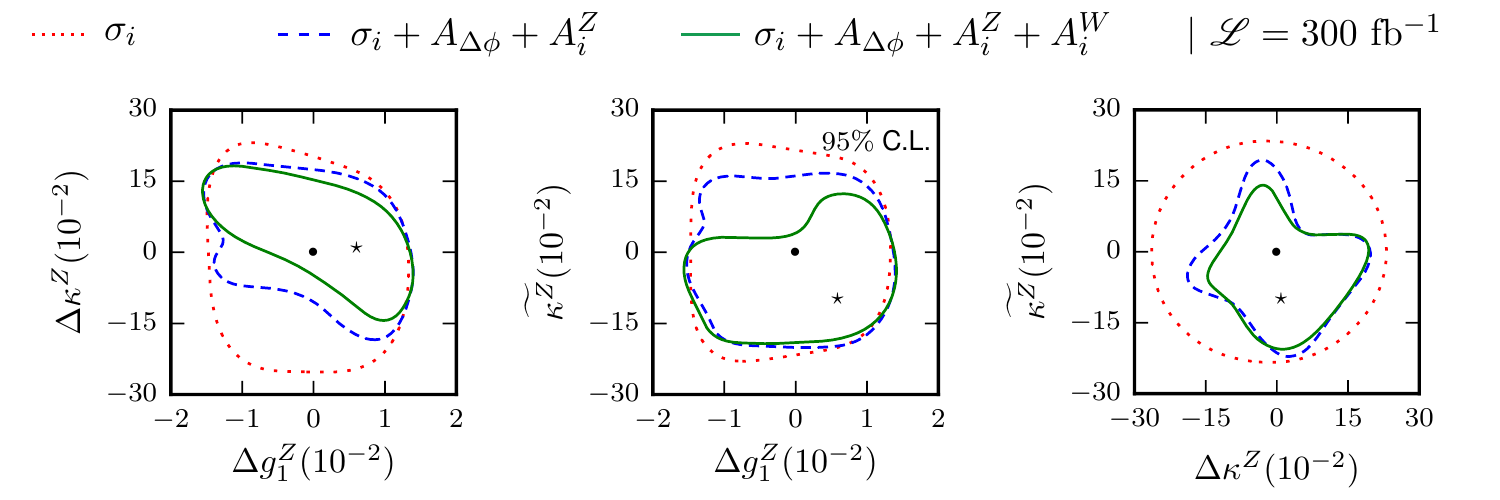}
	\caption{\label{fig:mcmcBench-Lum300fb}
		The marginalised  $2D$ projections at $95~\%$ BCI on 
		$\Delta g_1^Z$--$\Delta\kappa^Z$,    
		$\Delta g_1^Z$--$\wtil{\kappa^Z}$, and $\Delta\kappa^Z$--$\wtil{\kappa^Z}$ planes  from the MCMC with observables $\sigma_i$ ({\em dotted}/red line), $\sigma_i + A_{\Delta\phi}+A_i^Z$ 
		({\em dashed}/blue line) and $\sigma_i + A_{\Delta\phi}+A_i^Z+A_i^W$ ({\em solid}/green line)
		for {\tt aTGC-Bench} couplings $\{\Delta g_1^Z,\lambda^Z,\Delta\kappa^Z,\wtil{\lambda^Z},\wtil{\kappa^Z}\}=\{0.6,0.6,0.8,0.4,-10\}\times 10^{-2}$
		at $\sqrt{s}=13$ TeV and  ${\cal L}=300$ fb$^{-1}$.}
\end{figure}
\begin{figure}[h]
	\centering
\includegraphics[width=1.0\textwidth]{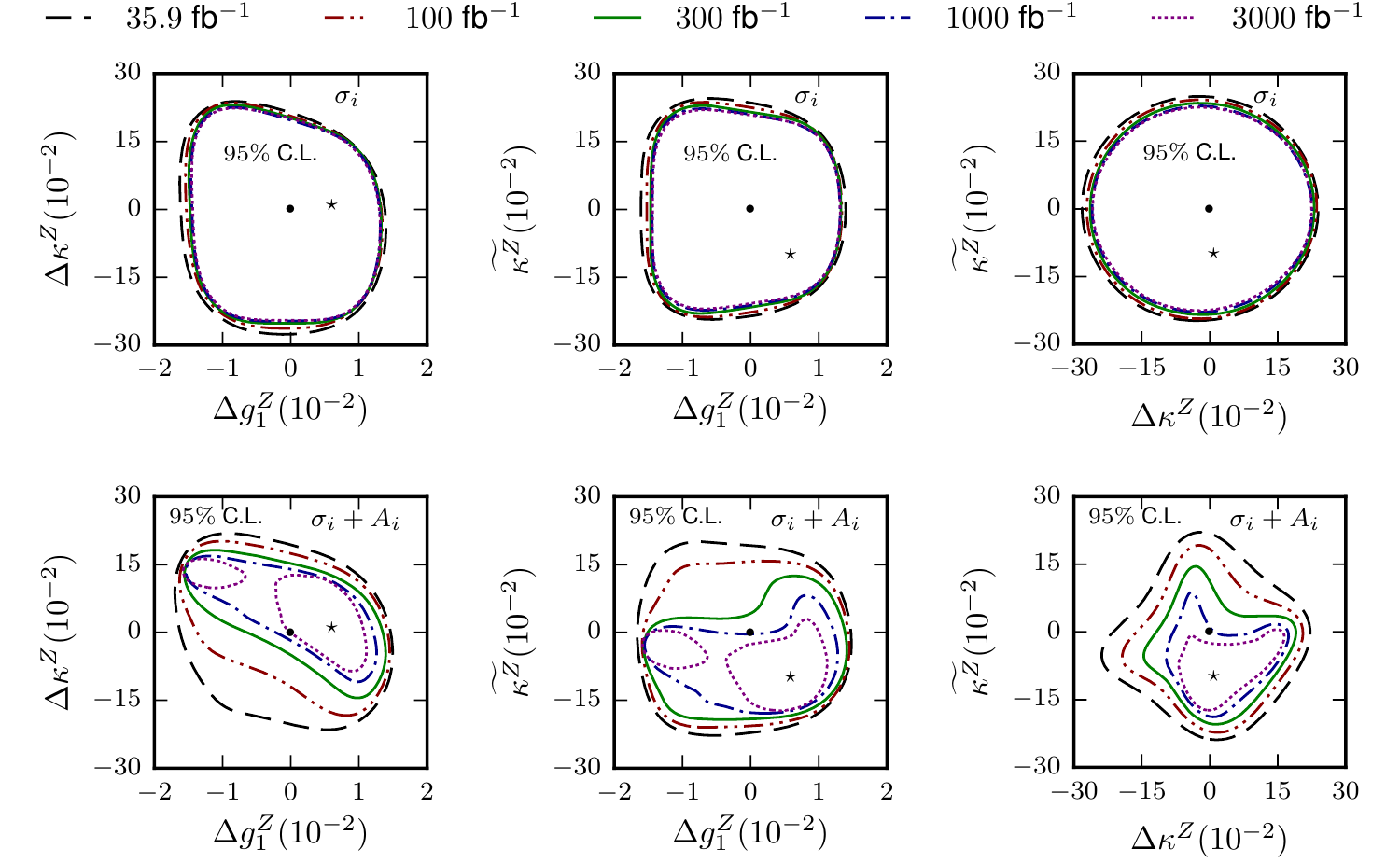}
	\caption{\label{fig:mcmcBench-VarLum}
		The marginalised  $2D$ projections at $95~\%$ BCI on 
		$\Delta g_1^Z$--$\Delta\kappa^Z$,  $\Delta g_1^Z$--$\wtil{\kappa^Z}$, and $\Delta\kappa^Z$--$\wtil{\kappa^Z}$ planes  from the MCMC with observables $\sigma_i$ in {\em top-row} and  $\sigma_i + A_{\Delta\phi}+A_i^Z+A_i^W$ in {\em bottom-row} for integrated  luminosities $35.9$ fb$^{-1}$ (outermost contours), $100$ fb$^{-1}$, $300$ fb$^{-1}$, $1000$ fb$^{-1}$, and  $3000$ fb$^{-1}$ (innermost contours) for {\tt aTGC-Bench} couplings $\{\Delta g_1^Z,\lambda^Z,\Delta\kappa^Z,\wtil{\lambda^Z},\wtil{\kappa^Z}\}=\{0.6,0.6,0.8,0.4,-10\}\times 10^{-2}$
		at $\sqrt{s}=13$ TeV.}
\end{figure}

The asymmetries are sub-dominant in constraining the couplings much like seen in Ref.~\cite{Rahaman:2018ujg}
for $pp\to ZZ$ case. 
But the asymmetries have a sense of directionality in the parameter space. To see this,  
we perform a toy analysis to extract {\em non-zero} anomalous couplings 
with pseudo data generated for the set of anomalous couplings of
\begin{equation}\label{eq:bench-aTGC}
 \text{{\tt aTGC-Bench}}:~ \{\Delta g_1^Z,\lambda^Z,\Delta\kappa^Z,\wtil{\lambda^Z},\wtil{\kappa^Z}\}=\{0.6,0.6,0.8,0.4,-10\}\times 10^{-2}
\end{equation}
using the MCMC method. These couplings are chosen to be within the current limits; see Table~\ref{tab:aTGC_constrain_form_collider}.  In Fig.~\ref{fig:mcmcBench-Lum300fb}, 
we show the posterior marginalised  $2D$ 
projections at $95~\%$ BCI on $\Delta g_1^Z$--$\Delta\kappa^Z$,  $\Delta g_1^Z$--$\wtil{\kappa^Z}$, and $\Delta\kappa^Z$--$\wtil{\kappa^Z}$ planes.  We draw the contours using   $\sigma_i$ only  ({\em dotted}/red line),  using $\sigma_i$ along with  
$A_{\Delta\phi}+A_i^Z$ ({\em dashed}/blue line) and all observables $\sigma_i + A_{\Delta\phi}+A_i^Z+A_i^W$ 
({\em solid}/green line)  for integrated luminosity of  ${\cal L}=300$ fb$^{-1}$.  The  dot ($\bullet$) points in the $2D$ contours represent the SM point, while  the 
 star-marks ($\star$)  represent the 
  couplings from {\tt aTGC-Bench}. As the asymmetries $A_{\Delta\phi}$ and  asymmetries of $Z$ ($A_i^Z$) 
  are added on top of the cross sections, the measurement gets better and it  improves
  further  when the asymmetries 
of $W$ ($A_i^W$) are added. The 
cross sections are blind to the orientation  of  {\tt aTGC-Bench} couplings and sensitive only to the magnitude 
of deviation from the SM. 
 The asymmetries, however, give direction to the measurement, e.g., in 
$\Delta\kappa^Z$--$\wtil{\kappa^Z}$ plane, $\sigma_i + A_{\Delta\phi}+A_i^Z$ give  tight and directional constraints. The above three planes  are shown again  in Fig.~\ref{fig:mcmcBench-VarLum} for varying luminosities of $35.9$ fb$^{-1}$ (outermost contours), $100$ fb$^{-1}$, $300$ fb$^{-1}$, $1000$ fb$^{-1}$, and  $3000$ fb$^{-1}$ (innermost contours) for observables $\sigma_i$ in {\em top-row} and  $\sigma_i + A_{\Delta\phi}+A_i^Z$ in {\em bottom-row}.
%For higher luminosities, the contours become tighter and centerd around the {\tt aTGC-Bench} couplings when using $\sigma_i+A_i$ ({\em bottom-row}), while  $\sigma_i$ alone ({\em top-row})  remain blind to the direction of  {\tt aTGC-Bench}. 
For higher luminosities, the $\sigma_i$ alone ({\em top-row}) do not yield improved
limits nor do they give any hint towards the direction of {\tt aTGC-Bench}. But the inclusion
of asymmetries $\sigma_i+A_i$ ({\em bottom-row}) give increasingly accurate determination of
the {\tt aTGC-Bench} points with increasing luminosity.
Thus,  this toy analysis indicates that the asymmetries would help  in the measurement of anomalous couplings at  high-luminosity  provided an  excess of events is observed at the LHC, and we interpret the deviation in terms of aTGC.

We note  that the   $3l+\cancel{E}_T$  excess   in the lower $p_T(Z)$ region at the LHC~\cite{Sirunyan:2019bez}     
interpreted by two extra scaler\cite{vonBuddenbrock:2019ajh} may be fitted by aTGC, which is beyond the scope of this present work.

%%%%%%%%%%%%%%%%%%%%%%%%%%%%%%%%%%%%%%%%%%%%%%%%%%%%%%%%%%%%%%%%%%%%%%%%%%%%%%%%%%%%%%%%%%%%%%%%%%%%%%%%%%%%%%%
\section{Conclusion}\label{sec:conclusion}
To conclude, we studied the $WWZ$ anomalous  couplings in the $ZW^\pm$ production at the LHC and 
examined the role of polarization asymmetries together with $\Delta\phi(l_W,Z)$ asymmetry and forward-backward asymmetry
on the estimation of limits on the anomalous couplings.
We reconstructed the missing neutrino momentum by choosing the small $|p_z(\nu)|$ from the two-fold solutions and estimated the $W$ polarization asymmetries, while the 
$Z$ polarization asymmetries are kept free from any reconstruction ambiguity.
We generated  events at NLO in QCD in {\tt mg5\_aMC}  for about $100$ sets of anomalous couplings and used them for the numerical fitting
of semi-analytic  expressions of all the observables as a function of the couplings. We estimated simultaneous limits on the
anomalous couplings using the MCMC method  for both the effective vertex formalism and the effective operator approach
for luminosities $35.9$ fb$^{-1}$, $100$ fb$^{-1}$, $300$ fb$^{-1}$, $1000$ fb$^{-1}$, and $3000$ fb$^{-1}$. The limits obtained  for ${\cal L}=35.9$ fb$^{-1}$   are tighter
than the limits available at the LHC (see Table~\ref{tab:aTGC_constrain_form_collider} \&~\ref{tab:simul-limits-Lag-Op-OpLag}) except on $c_W$ (and $\Delta\kappa^Z$).  
The asymmetries are helpful in extracting the values of anomalous
couplings  if a deviation from the SM is observed at the LHC. We performed a toy analysis of parameter extraction with some benchmark
aTGC couplings and observed that the inclusion of  asymmetries to  the cross sections improves the parameter extraction significantly at high-luminosity.

%%%%%%%%%%%%%%%%%%%%%%%%%%%%%%%%%%%%%%%%%%%%%%%%%%%%%%%%%%%%%%%%%%%%%%%%%%%%%%%%%%%%%%
%%%%%%%%%%%%%%%%%%%%%%%%%%%%%%%%%%%%%%%%

\vspace{0.5 cm}
\noindent \textbf{Acknowledgements:} R.R. thanks Department of Science 
and Technology, Government of India for support through DST-INSPIRE Fellowship 
for doctoral program, INSPIRE CODE IF140075, 2014. 
RKS acknowledges SERB, DST, Government of India through the project EMR/2017/002778.
The authors thank the anonymous referee for his/her suggestions for the improvement of this article.

%%%%%%%%%%%%%%%%%%%%%%%%%%%%%%%%%%%%%%%%%%%%%%%%%%%%%%%%%%%%%%%%%%%%%%%%%%%%%%%%%%%%%%%%%%%%%%%%%%%
\appendix
\section{Fitting procedure in obtaining observables as a function of couplings}\label{app:fitting}
The SM+aTGC  events are generated for about $100$ set  of couplings
$$\{c_i\}=\{\Delta g_1^Z,\lambda^Z,\Delta\kappa^Z,\wtil{\lambda^Z},\wtil{\kappa^Z}\}$$
in both processes.  The values of all the observables are obtained for  the set couplings in the optimized
cuts (Table~\ref{tab:m3lpTZcut-on-Asym}), and then those are used for numerical fitting to obtain the semi-analytical expression of all the observables as a function of the couplings.
For the cross sections the following $CP$-even expression is used to fit the data:
\begin{equation}\label{eq:sigma-fit}
\sigma(\{c_i\}) =\sigma_{SM} + \sum_{i=1}^{3} c_i \times \sigma_i +\sum_{i=1}^{5} (c_i)^2 \times \sigma_{ii}  + \frac{1}{2}\sum_{i=1}^{3}\sum_{j(\ne i)=1}^{3} c_i c_j \times \sigma_{ij}
+ c_4 c_5 \times \sigma_{45} \ \ .
\end{equation}
For asymmetries, the numerator and the denominator  are fitted separately and then used as
\begin{equation}
A_j(\{c_i\})=\dfrac{\Delta\sigma_{A_j}(\{c_i\})}{\sigma_{A_j}(\{c_i\})} \ \  .
\end{equation}
The numerator ($\Delta\sigma_A$) of $CP$-odd asymmetries are fitted with  the $CP$-odd expression
\begin{equation}\label{eq:CP-odd-fit}
\Delta\sigma_A(\{c_i\}) =\sum_{i=4}^{5} c_i \times \sigma_i + \sum_{i=1}^{3} \left( c_i c_4 \times\sigma_{i4}
+  c_i c_5 \times\sigma_{i5}\right) \ \ .
\end{equation}
The denominator ($\sigma_{A_j}$) of all the asymmetries and the numerator  ($\Delta\sigma_A$) of $CP$-even
asymmetries are fitted with the $CP$-even expression given in Eq.~(\ref{eq:sigma-fit}).

\begin{figure}
	\centering
	\includegraphics[width=0.496\textwidth]{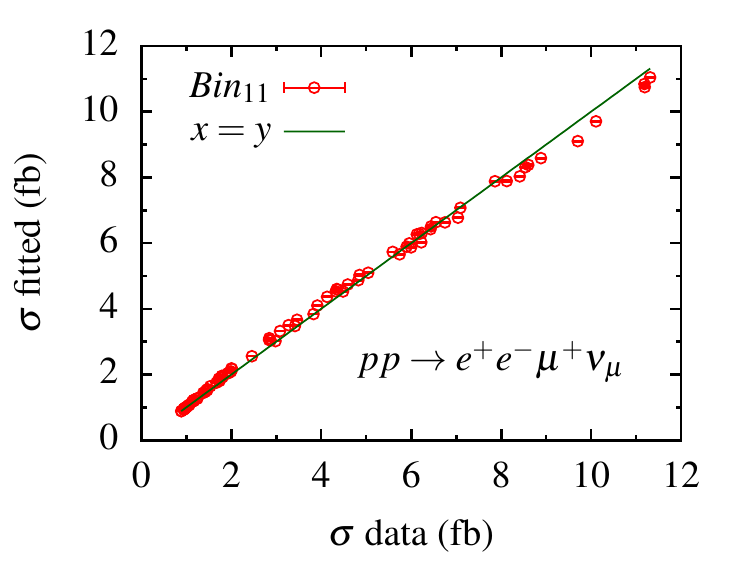}
	\includegraphics[width=0.496\textwidth]{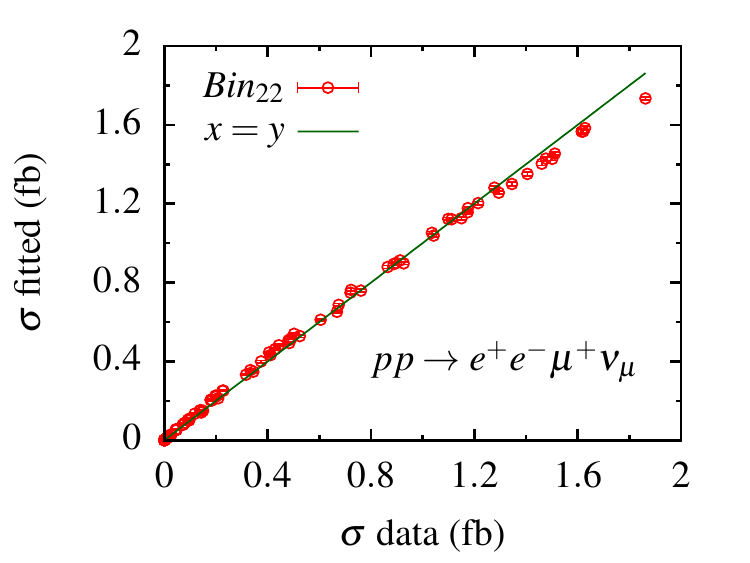}
	\includegraphics[width=0.496\textwidth]{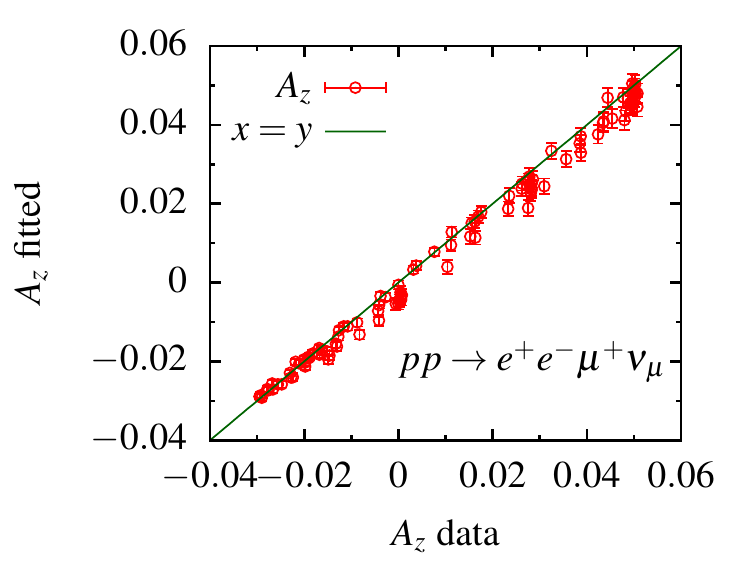}
	\includegraphics[width=0.496\textwidth]{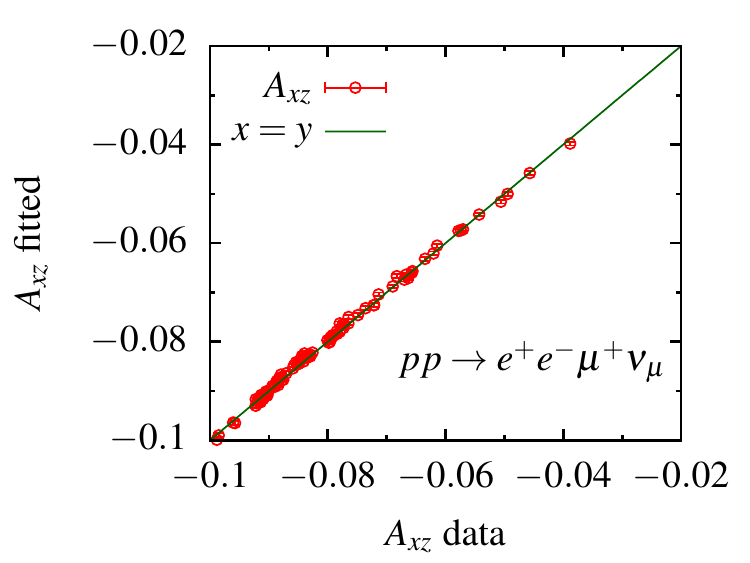}
	\caption{\label{fig:Fitt-Test-sigma-ZWp}
The simulated data (in $x$-axis) vs. fitted values (in $y$-axis) for the cross section in the two diagonal bins ({\em top-panel})
 and the polarization asymmetries $A_z$ and $A_{xz}$ ({\em bottom-panel}) in
in $ZW^+$ production in $e^+e^-\mu^+\nu_\mu$ channel at the LHC at $\sqrt{s}=13$ TeV. 
}
\end{figure}
We use the MCMC method to
fit the coefficients  of the cross sections with positivity demand, i.e., $\sigma(\{c_i\})\ge 0$.
We use $80~\%$ data to fit the coefficients of the cross sections,  and then the fitted expressions are
validated against the rest   $20~\%$ of the  data  and found to be matching within $2\sigma$ MC error. 
We generated  $10^7$ events to keep the MC  error as small as possible,  even in the tightest
optimized cuts. For example, the $A_{zz}$ in $ZW^+$ has the tightest cut on $m_{3l}$ (see Table~\ref{tab:m3lpTZcut-on-Asym}) and yet have very small ($0.2~\%$) MC error (see Table~\ref{tab:SM-values-Asym-Pol}).  
In Fig.~\ref{fig:Fitt-Test-sigma-ZWp} fitted values of observables are compared against the simulated data for
the cross section in two diagonal bins ({\em top-panel})
and the polarization asymmetries $A_z$ and $A_{xz}$ ({\em bottom-panel}) 
in $ZW^+$ production in $e^+e^-\mu^+\nu_\mu$ channel as representative. The fitted values seem
to  agree with the simulated data used within the MC error.
%%%%%%%%%%%%%%%%%%
\section{Standard Model values of the asymmetries and polarizations}\label{app:SM-values-Asym}
In Table~\ref{tab:SM-values-Asym-Pol}, we show the SM estimates (with $1\sigma$ MC error) of the polarization 
asymmetries of $Z$ and $W$ and their corresponding polarizations along with the other 
asymmetries for our selection cuts ({\tt sel.cut}) given in Eq.~(\ref{eq:selection-cuts}) 
and optimized cuts ({\tt opt.cut}) given Table~\ref{tab:m3lpTZcut-on-Asym}.  
A  number of events  of $N\simeq9.9\times 10^6$ satisfy our selection cuts,  which give the same error
($\delta A_i=1/\sqrt{N}$)  for all the asymmetries, and hence  they are given in the top row. As the
optimized cuts for $W$ are the same for all the asymmetries, the errors for them are also given in the top row. For the
optimized cuts of $Z$ observables, however, the number of events varies, and hence the MC errors are given to each asymmetries.  
The $CP$-odd polarizations $p_y$, $T_{xy}$, $T_{yz}$, 
and their corresponding asymmetries are consistent with zero in the SM within MC error.
%\begin{table}
\afterpage{\clearpage
\begin{sidewaystable}
\caption{\label{tab:SM-values-Asym-Pol} The SM values with $1\sigma$ MC error of the polarization 
asymmetries of $Z$ and $W$ and their corresponding polarizations along with the other 
asymmetries in  $ZW^\pm$ production in the $e^+e^-\mu^\pm+\cancel{E}_T$ channel are shown 
for event selection cuts ({\tt sel.cut}) given in 
Eq.~(\ref{eq:selection-cuts}) and optimized cuts ({\tt opt.cut}) given Table~\ref{tab:m3lpTZcut-on-Asym}. 
%For the polarization we use $\alpha_Z=(R_l^2-L_f^2)/(R_l^2+L_f^2)=-0.219325$, $\alpha_W=-1$. 
 %Number of events for {\tt sel.cut} $Nt(ZW^+)=9985450(\delta A = 0.0003)$ and $Nt(ZW^-)=9987560(\delta A = 0.0003)$.  
 }
\renewcommand{\arraystretch}{1.50}
\centering
\begin{scriptsize}
\begin{tabular*}{\textwidth}{@{\extracolsep{\fill}}|l|l|l|l|l|l|l|l|l|@{}}\hline
& \multicolumn{4}{c|}{$ZW^+$}& \multicolumn{4}{c|}{$ZW^-$} \\ \hline
&  \multicolumn{2}{c|}{$Z$}& \multicolumn{2}{c|}{$W^+$} &  \multicolumn{2}{c|}{$Z$}& \multicolumn{2}{c|}{$W^-$}\\ \hline
${\cal O}$& {\tt sel.cut} & {\tt opt.cut}& {\tt sel.cut} & {\tt opt.cut}& {\tt sel.cut} & {\tt opt.cut}& {\tt sel.cut} & {\tt opt.cut}\\ 
$\delta A_i$ &$\pm 0.0003$&          &$\pm 0.0003$&$\pm 0.0007$ &$\pm 0.0003$ &  &$\pm 0.0003$ &$\pm 0.0007$ \\ \hline
$A_x$          & $-0.0196$&$-0.0150\pm 0.0008$ &$-0.2303$ &$-0.0550$&$+0.0074$&$-0.0046\pm 0.0010$&$-0.0826$&$-0.0001$  \\ 
$p_x$          & $+0.1192\pm0.0018$&$+0.0912\pm0.0049$ &$+0.3071\pm0.0004$ &$0.0733\pm0.0009$&$-0.0450\pm0.0018$&$+0.0280\pm0.0061$&$+0.110\pm0.00041$&$+0.00013\pm0.0009$  \\ \hline
$A_y$          &$+0.0003$ &$+0.0004\pm 0.0007$ &$-0.0007 $ &$-0.0005$&$-0.0013$&$-0.0021\pm0.0007$&$0.0$&$+0.0007$  \\ 
$p_y$          &$-0.0018\pm0.0018$ &$-0.0024\pm0.0146$ &$+0.0009\pm0.0004$ &$+0.0006\pm0.0009$&$+0.0079\pm0.0018$&$+0.0127\pm0.0042$&$0.0\pm0.0004$&$-0.0009\pm0.0009$  \\ \hline
$A_z$          &$-0.0040$ &$+0.0502\pm 0.0025$ &$ +0.1337$ &$+0.6615$&$+0.0316$&$+0.0482\pm0.0019$&$+0.1954$&$+0.7381$  \\ 
$p_z$          &$+0.0243\pm0.0018$ &$-0.3051\pm0.0152$ &$-0.1783\pm0.0004$ &$-0.8820\pm0.0009$&$-0.1921\pm0.0018$&$-0.2930\pm0.0115$&$-0.2605\pm0.0004$&$-0.9841\pm0.0009$  \\ \hline
$A_{xy}$       &$-0.0017$ &$+0.0005\pm 0.0007$ &$-0.0011$ &$-0.0006$&$+0.0008$&$+0.0014\pm0.0007$&$+0.0013$ &$-0.0003$ \\ 
$T_{xy}$       &$-0.0033\pm0.0006$ &$+0.00096\pm0.0013$ &$-0.0021\pm0.0006$ &$-0.0012\pm0.0013$&$+0.0015\pm0.0006$&$+0.0027\pm0.0013$&$+0.0025\pm0.0006$ &$-0.0006\pm0.0013$ \\ \hline
$A_{xz}$       &$+0.0196$ &$+0.0914\pm 0.0004$ &$+0.0048$ &$-0.0063$&$+0.0961$&$+0.0547\pm0.0006$&$+0.0010$ &$-0.0136$ \\ 
$T_{xz}$       &$+0.0377\pm0.0006$ &$+0.1758\pm0.0008$ &$+0.0092\pm0.0006$ &$-0.0121\pm0.0013$&$+0.1849\pm0.0006$&$+0.1052\pm0.0011$&$+0.0019\pm0.0006$ &$-0.0262\pm0.0013$ \\ \hline
$A_{yz}$       &$+0.0002$ &$-0.0001\pm 0.0004$ &$+0.0003$ &$-0.0005$&$-0.0017$&$-0.0016\pm0.0003$&$+0.0001$&$-0.0001$  \\ 
$T_{yz}$       &$+0.0004\pm0.0006$ &$-0.0002\pm0.0008$ &$+0.0006\pm0.0006$ &$-0.0009\pm0.0013$&$-0.0033\pm0.0006$&$-0.0031\pm0.0006$&$+0.0002\pm0.0006$&$-0.0002\pm0.0013$  \\ \hline
$A_{x^2-y^2}$  &$-0.0878$ &$-0.0925\pm 0.0019$ &$-0.0266$ &$-0.1326$&$-0.0935$&$-0.0899\pm0.0012$&$-0.0923$&$-0.1588$  \\ 
$T_{xx}-T_{yy}$&$-0.3378\pm0.0011$ &$-0.3559\pm0.0073$ &$-0.1023\pm0.0011$ &$-0.5102\pm0.0027$&$-0.3597\pm0.0011$&$-0.3459\pm0.0046$&$-0.3551\pm0.0011$&$-0.6110\pm0.0027$  \\ \hline
$A_{zz}$       & $-0.0137$&$+0.0982\pm 0.0024$ &$+0.0519$ &$+0.1406$&$+0.0030$&$+0.0863\pm0.0048$&$+0.1046$&$+0.2547$  \\ 
$T_{zz}$       & $-0.0298\pm0.0006$&$+0.2138\pm.0052$ &$+0.1130\pm0.0006$ &$+0.3061\pm0.0015 $&$+0.0065\pm0.0006$&$+0.1879\pm0.0104$&$+0.2277\pm0.0006$&$+0.5546\pm0.0015$  \\ \hline
$A_{fb}$       &$+0.6829$ &$+0.4475\pm 0.0009$  &  $+0.4699$&$+0.2627$ &$+0.6696$&$+0.2791\pm0.0025$&$+0.2060$&$+0.3174$ \\ \hline\hline
%$A_{fb}$       &$+0.6829$ &$+0.4475\pm 0.0009$  &  ---&--- &$+0.6696$&$+0.2791\pm0.0025$&---&--- \\ \hline\hline
&  \multicolumn{2}{c|}{{\tt sel.cut} }& \multicolumn{2}{c|}{{\tt opt.cut}} &  \multicolumn{2}{c|}{{\tt sel.cut} }& \multicolumn{2}{c|}{{\tt opt.cut}}\\ \hline
$A_{\Delta\phi}$&  \multicolumn{2}{c|}{$-0.3756\pm 0.0003$}& \multicolumn{2}{c|}{$-0.4151\pm 0.0022$} &  \multicolumn{2}{c|}{$-0.3880\pm 0.0003$}& \multicolumn{2}{c|}{$-0.4208\pm 0.0025$}\\ \hline
\end{tabular*}
\end{scriptsize}
\end{sidewaystable}
%\end{table}
\clearpage }

%%%%%%%%%%%%%%%%%%%%%%%%%%%%%%%%%%%%%
\bibliography{CITATIONS-WZ}
\bibliographystyle{utphysM}
%%%%%%%%%%%%%%%%%%%%%%%%%%%%%%%%%%%%%%
\end{document}